\documentclass{aa} 

\usepackage{graphicx}
\usepackage{txfonts}
\usepackage{hyperref}
\usepackage{bm}

\newcommand\id{{\mathrm d}}

\newcommand\bx{{\bm x}}
\newcommand\by{{\bm y}}

\newcommand\bk{{\bm k}}

\newcommand\bu{{\bm u}}

\newcommand\bK{{\bm K}}

\newcommand\Cref{C_{\text{ref}}}
\newcommand\cref{\Cref}

\newcommand\cK{\mathcal{K}}

\newcommand\PFFT{P^{\rm{fft}}}
\newcommand\tauNR{\tau_{\rm{NR}}}

\DeclareMathOperator{\floor}{floor}

\usepackage{hyphenat}
\defcitealias{Hanasoge2012}{HDS2012}

\begin{document} 

\title{Helioseismological determination of the subsurface spatial spectrum of solar convection: Demonstration using numerical simulations}

\titlerunning{Helioseismological determination of the subsurface velocity spectrum of solar convection}

\author{
        Vincent G. A. Böning\inst{1}
        \and
        Aaron C. Birch\inst{1}
        \and
        Laurent Gizon\inst{1,2,3}
        \and
        Thomas L. Duvall, Jr.\inst{1}
}

\institute{
        {Max-Planck-Institut f\"ur Sonnensystemforschung, Justus-von-Liebig-Weg 3, 37077 G\"ottingen, Germany}\\
        \email{boening@mps.mpg.de}
        \and
        {Institut f\"ur Astrophysik, Georg-August-Universit\"at G\"ottingen, Friedrich-Hund-Platz 1, 37077 G\"ottingen, Germany}
        \and
        {Center for Space Science, NYUAD Institute, New York University Abu Dhabi, PO Box 129188, Abu Dhabi, UAE}
        }

\authorrunning{Böning et. al.}

\date{Received ???; Accepted ???}


\abstract
{Understanding convection is important {in} stellar physics, for example, when it is an input in stellar evolution models. Helioseismic estimates of convective flow amplitudes {in deeper regions of the solar interior} disagree by orders of magnitude {among themselves and with} simulations.}
{We aim to assess the validity of an existing upper limit of {solar} convective flow amplitudes {at a depth of 0.96 solar radii} obtained using time-distance helioseismology and several simplifying assumptions.}
{We generated synthetic observations for convective flow fields from a magnetohydrodynamic simulation (MURaM) using travel-time sensitivity functions and a noise model. We compared the estimates of the flow amplitude with the actual value of the flow.}
{{For the scales of interest ($\ell<100$), w}e find that the current procedure for obtaining an upper limit gives the correct order of magnitude of the flow for the given flow fields. We also show that this estimate is not an upper limit in a strict sense because it underestimates the flow amplitude at the largest scales by a factor of about two because the scale dependence of the signal-to-noise ratio has to be taken into account. {After correcting for this and after taking the dependence of the measurements on direction in Fourier space into account, we show that the obtained estimate is indeed an upper limit.}}
{We conclude that time-distance helioseismology is able to correctly estimate the order of magnitude (or an upper limit) of solar convective flows {in the deeper interior} when {the} vertical correlation {function} {of the different flow components} {is known} {and the scale dependence of the signal-to-noise ratio is taken into account}. {We suggest that future work should include information from different target depths to better separate the effect of near-surface flows from those at greater depths. In addition, the measurements are sensitive to all three flow directions, which should be taken into account.}}


\maketitle


%

\section{Introduction}
\label{secIntro}

{Helioseismic inferences of convective motions in the solar interior yield consistent results in near-surface layers using different techniques \citep[e.g.,][]{DeRosa2000,Braun2003,Hindman2004,GB2004,Langfellner2014,Langfellner2015}. One important statistical property of convective motions is their spatial power spectrum, which quantifies the scale-dependent amplitude of the velocities. Inferences of the power spectrum of convective motions in deeper regions of the solar interior}  have led to conclusions that differ by more than an order of magnitude {(\citealp{Hanasoge2012}, hereafter \citetalias{Hanasoge2012}; \citealp{Greer2015})}. Understanding these differences and obtaining reliable estimates of the spectrum of convective motions in the Sun is important for several processes in stellar interiors, such as {the maintenance of} differential rotation, and {for understanding the role of convection in the emergence of active regions.} This is even more urgent because convective velocities from numerical simulations of solar convection exhibit large-scale convective motions that are far too strong, even compared to surface observations \citep[]{Gizon2012PNASconvection,Lord2014}. {This last issue is often referred to as the convective conundrum.}

To date, the only methods that have been used to constrain the spectrum of convective motions in the {deeper} solar interior are time-distance helioseismology \citep{Hanasoge2010,Hanasoge2012} and ring-diagram analysis \citep[][]{Greer2015}. {As helioseismic techniques, both infer solar interior flows from surface observations of stochastically excited sound waves and surface-gravity waves, which propagate through the interior. To do so, time-distance helioseismology \citep{Duvall1993} uses travel times of waves that travel different distances and hence probe different depths. In a ring-diagram analysis \citep{Hill1988}, observations of the local power spectrum of the surface oscillations are used and the Doppler-shifted frequency of the seismic waves is measured by way of a background flow field. \citet{GB2005} and \citet[][]{Gizon2010} reviewed these local helioseismic techniques, and \citet{Hanasoge2016citavi} reviewed their application to probing interior convection. In addition, global helioseismology can be used to infer properties of time-dependent flows such as convection \citep[e.g.,][]{Roth2003,Woodard2016,Mani2020}. 
This method relies on global properties of the oscillations such as mode frequencies or amplitudes. Time-dependent flows result in mode coupling, which can be measured in global observations and be used to reveal flow properties.

}

{For the case of time-distance helioseismology}, \cite{Hanasoge2010,Hanasoge2012} estimated an upper limit of convective motions based on several simplifying assumptions, notably that the signal-to-noise ratio {of the helioseismic measurements} is independent of the {horizontal spatial} scale and that the signal in travel-time measurements is predominantly due to the flow in one direction at or near a targeted depth.

We here therefore aim to assess the validity of the method suggested by {\citetalias{Hanasoge2012}} to infer an upper limit on convective flow amplitudes in the solar interior. To do this, we generated a number of realizations of synthetic travel-time maps {using background flow fields} from {existing} numerical magnetohydrodynamic (MHD) simulation{s of quiet-Sun convection} {\citep{Lord2014}}. As we know the input flow, we can assess the performance of the resulting estimate and the underlying assumptions. {To generate the synthetic observations, we {employed a forward model of the signal and a noise model. In the forward model, the signal in the travel-time maps is computed using the first} Born approximation \citep[][]{Boening2016}}{, which takes single-scattering perturbations to the wave field into account \citep{GB2002}. We obtain realizations of the stochastic noise from a Gaussian noise model in Fourier space \citep{GB2004,Fournier2014}, which is based on the assumption of a large number of random wave sources homogeneously distributed in space and time. These are well-tested methods used in the community \citep[e.g.,][]{Boening2017Inversions,Gizon2020}.}

\section{Methods}

We consider the problem of estimating subsurface convective flow amplitudes {from simulations} in Cartesian geometry. We assume that the {statistical properties of the} flow {do not depend on location within} a horizontal layer.

{Clearly, the use of Cartesian geometry is a simplification, especially when the interest is on larger horizontal scales. The main effects of this simplification are that (i) the resolution in Fourier space is reduced compared to covering the entire sphere, and (ii) effects of the spherical geometry such as rotation and curvature on the flow are not modeled. These are important shortcomings that should be addressed in future studies. However, given the order-of-magnitude difference between \citetalias{Hanasoge2012} and \citet{Greer2015}, our main interest in this work is to verify whether one of the proposed methods \citepalias{Hanasoge2012} is working at all, and potentially, to give clues on the cause of the differences.}

{In addition, the essential problem that convective flow amplitudes are larger in simulations than in observations is also present for Cartesian simulations of convective flows \citep[][see the comparison to surface observations in Fig. 5]{Lord2014}. In Cartesian geometry, more realistic simulations of convection are available than in spherical geometry because radiative transfer can be better included and smaller turbulent scales can be resolved \citep{Kupka2017}.}

{In this section, we first summarize the methods that we used to model the method of \citetalias{Hanasoge2012} and to create synthetic data. We then summarize the assumptions that were made by \citetalias{Hanasoge2012} to obtain an upper limit on the spectrum of convection to be able to study the validity of these assumptions in the following sections.}

\subsection{Convective flow fields from MURaM}
\label{secMURaM}

{As input for generating synthetic observations, w}e used 66 consecutive snapshots of MHD simulations of solar convection. {These simulations were obtained by \cite{Lord2014thesis} using the MURaM code} \citep{Vogler2005} {with the vertical magnetic flux fixed to zero. The details for these simulations are given in \citet[][see the case ``zero net flux'' in Tables~2.1 and~5.1, as well as Sections~2 and~5]{Lord2014thesis}\footnote{{The Ph.D. thesis of \citet{Lord2014thesis} ``Deep Convection, Magnetism and Solar Supergranulation'' may be downloaded from 
\url{https://ui.adsabs.harvard.edu/abs/2014PhDT.......241L/abstract}.}}. Results for the purely hydrodynamic case of this simulation setup were reported in \citet{Lord2014}. After the hydrodynamic simulations had reached a statistically steady state, \cite{Lord2014thesis} initialized the magnetic field with a root mean square magnetic field strength of $0.5\,\rm{G}$ while keeping the  horizontally averaged vertical magnetic flux at zero. The dynamo-related magnetic field amplification resulted in a photospheric root mean square magnetic field strength of $140\,\rm{G}$ and a photospheric mean unsigned vertical magnetic flux of $37.2\,\rm{G}$.

\citet{Lord2014thesis} showed snapshots of the vertical velocity $u_z$ of the hydrodynamic case in Figures~2.2 and~2.3 and horizontally averaged density, temperature, and pressure profiles in Figure~2.5, which are close to a standard solar model \citep{JCD1996}. Snapshots and horizontal averages of background quantities are very similar for the magnetic case \citep[see][]{Lord2014thesis}. We show a snapshot of the horizontal velocity $u_x$ in Figure~\ref{figExampleUx}.}

{W}e used 66 consecutive snapshots {that were archived by \citet{Lord2014thesis}. They were} extracted at a cadence of 3.8~hours of solar time, and {rebinned} to $512\times512\times384$ grid points with a resolution of $384\times384\times128\,\rm{km}^3$ in a {Cartesian} box sized $196\times196\times50\,\rm{Mm}^3$ in the $(x,y,z)$ directions{, where $z$ points upward {and $L=196\,\rm{Mm}$ is the horizontal extent of the box}.} {The original spatial resolution had twice as many grid points in all directions. The reduction in resolution only changes the smaller scales in the simulation, while in this study, we are interested in the very largest scales, which are relatively unaffected by the reduction in resolution. The simulation extended to a depth of $z=-48.5\,\rm{Mm}$ ($r=0.93\,R_\sun$, or about 16 cumulative pressure scale heights from the surface). This study is targeted at flows around $r=0.96\,R_\sun$, which is about 2.2 local pressure scale heights from the bottom boundary.}

{The helioseismic data in \citetalias{Hanasoge2012} were averaged over time periods of one day. The data are thus sensitive to temporal averages of convective motions. We here mimicked this temporal averaging by subdividing} the simulated cubes into 11 chunks of $6\times3.8\,\rm{h}=22.8\,\rm{h}$ and by averaging the flow field over the six snapshots for each chunk. {This is further discussed in Sections~\ref{secSummaryAssumptions} and~\ref{secConlusions}.}

\subsection{Forward model in Cartesian geometry}

As the simulated flow fields are given in Cartesian geometry and are horizontally periodic, we adopt a Cartesian geometry for the rest of this paper. The Cartesian box was centered at the equator. In order to convert from Cartesian {coordinates  $(x,y,z)$} into spherical coordinates $(r,\theta,\phi)$ {and vice versa,} {where {$r$ is the distance from solar center,} $\theta$ is the colatitude and $\phi$ is the longitude}, we used the relations $x = \phi R_\sun, y = (\pi/2 - \theta) R_\sun, z = r -R_\sun$. To facilitate comparison with \citetalias{Hanasoge2012}, we continue to refer to the $z$ coordinate by the corresponding value for $r$.

In Cartesian geometry, the convolution theorem can be used, see Appendix~\ref{appGB04FFT}, to write the relation between travel times $\tau$ and flows $\bu=(u_x,u_y,u_z)${, which is mediated by the sensitivity function $\bK=(K_x,K_y,K_z)$} in the Fourier domain,
\begin{align}
\tau(\bx) &= \sum_z h_z \, \Bigg( h_x^2  \sum_{\bx'} \bu(\bx',z) \cdot \bK(\bx-\bx',z)  \Bigg)  + \epsilon(\bx), \label{eqtau}\\
\tau(\bk) &=  \sum_z h_z \, \Bigg( (2\pi)^2  \bu(\bk,z) \cdot \bK(\bk,z) \Bigg) + \epsilon(\bk), \label{eqtauKepsCart}
\end{align}
where we mark the {horizontal} Fourier transform only by the use of the {horizontal} Fourier variable {or wave vector,} $\bk$, see Appendix~\ref{appGB04FFT}{, and we used $\ell=k R_\sun$ to convert from {horizontal} wave number $k=|\bk|$ into harmonic degree.} {In Eq.~\eqref{eqtau}, we have summed over all vertical grid points $z$ and horizontal grid points $\bx'$, we multiplied by the uniform vertical grid spacing $h_z$ and horizontal grid spacing $h_x=h_y$, and we used $\bx - \bx'=(x-x',y-y')$ to define a horizontal convolution.} {{In our computations,} we used the discretization in Eq.~\eqref{eqtau}, but for brevity in notation, we write in the remainder of this paper {for the vertical integral}}
\begin{align}
\sum_z h_z \leftrightarrow \int \id z.
\end{align}

\begin{figure}
        \centering
        \includegraphics[width=\linewidth]{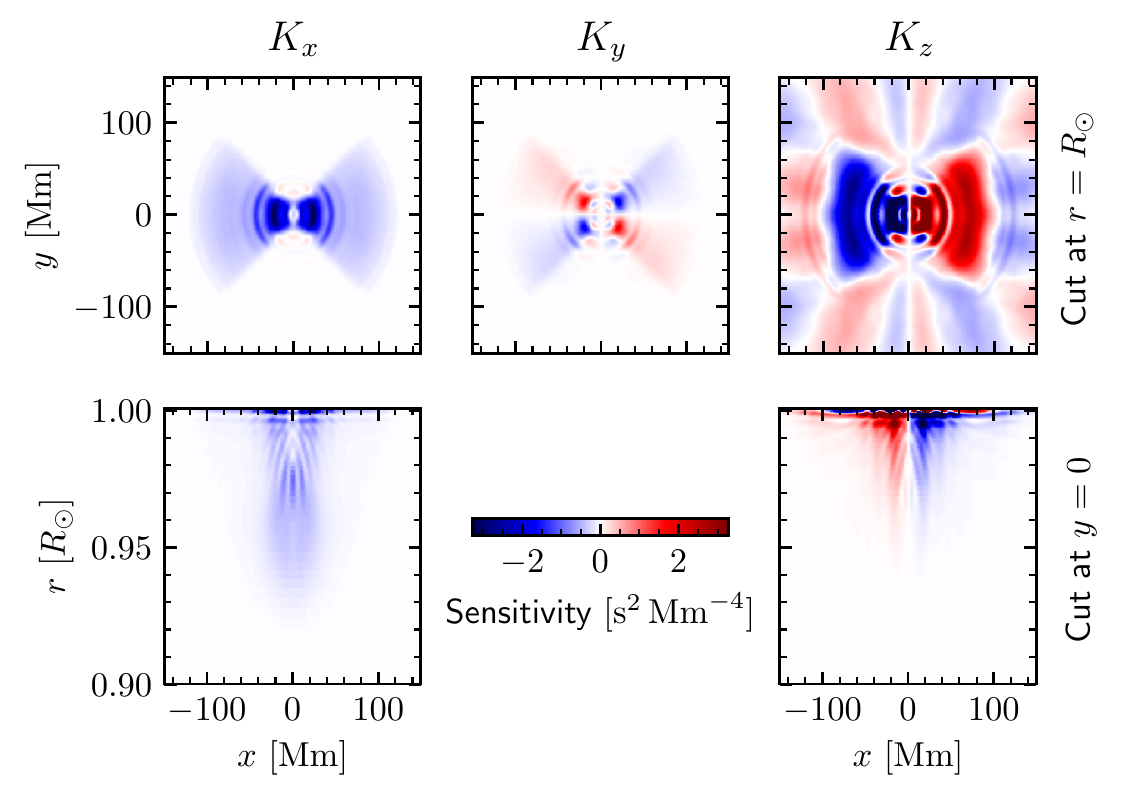}
        \caption{Sensitivity of deep-focus travel times (target depth $0.96 R_\sun$) to flows. The observational setup is taken from \citetalias{Hanasoge2012} {and is described in detail in Appendix~\ref{appGeometry}}. {We computed the sensitivity functions using the first Born approximation \citep[][]{Boening2016}.}}
        \label{figKernels}
\end{figure}

We here considered the deep-focus travel-time geometry used by \citetalias{Hanasoge2012} for the target depth of $r=0.96R_\sun$ and computed travel-time sensitivity kernels$\bK$  for this measurement using the Born approximation and the method of \cite{Boening2016}, see Figure~\ref{figKernels}. This method assumes that the flow is constant in time. {Sensitivity functions obtained using the Born approximation compare well in different approaches \citep[e.g.,][]{Birch2007,Burston2015,Boening2016,Mandal2017}. To compute the sensitivity functions, we used a spatial grid that was four times as wide as the MURaM grid to capture the entire sensitivity function. It was then Fourier-projected on the MURaM grid using the periodicity of the flow field. Furthermore, we used only half the spatial resolution for computing kernels. This is sufficient because the smallest seismic wavelength is still more then ten times larger than the spatial resolution.} {Appendix~\ref{appGeometry} gives further details on the deep-focus travel-time geometry{, which was modeled using the kernels. The travel-time geometry was designed by \citetalias{Hanasoge2012} to be predominantly sensitive to $u_x$ flows (the east-west component of the flow) close to a target depth of $0.96\,R_\sun$. Figure~\ref{figKernels} shows that the travel-time measurements are sensitive to flows in all spatial directions and especially to near-surface flows, although there is some degree of focussing for $K_x$.}}

\begin{figure}
        \centering

        \includegraphics[width=\linewidth]{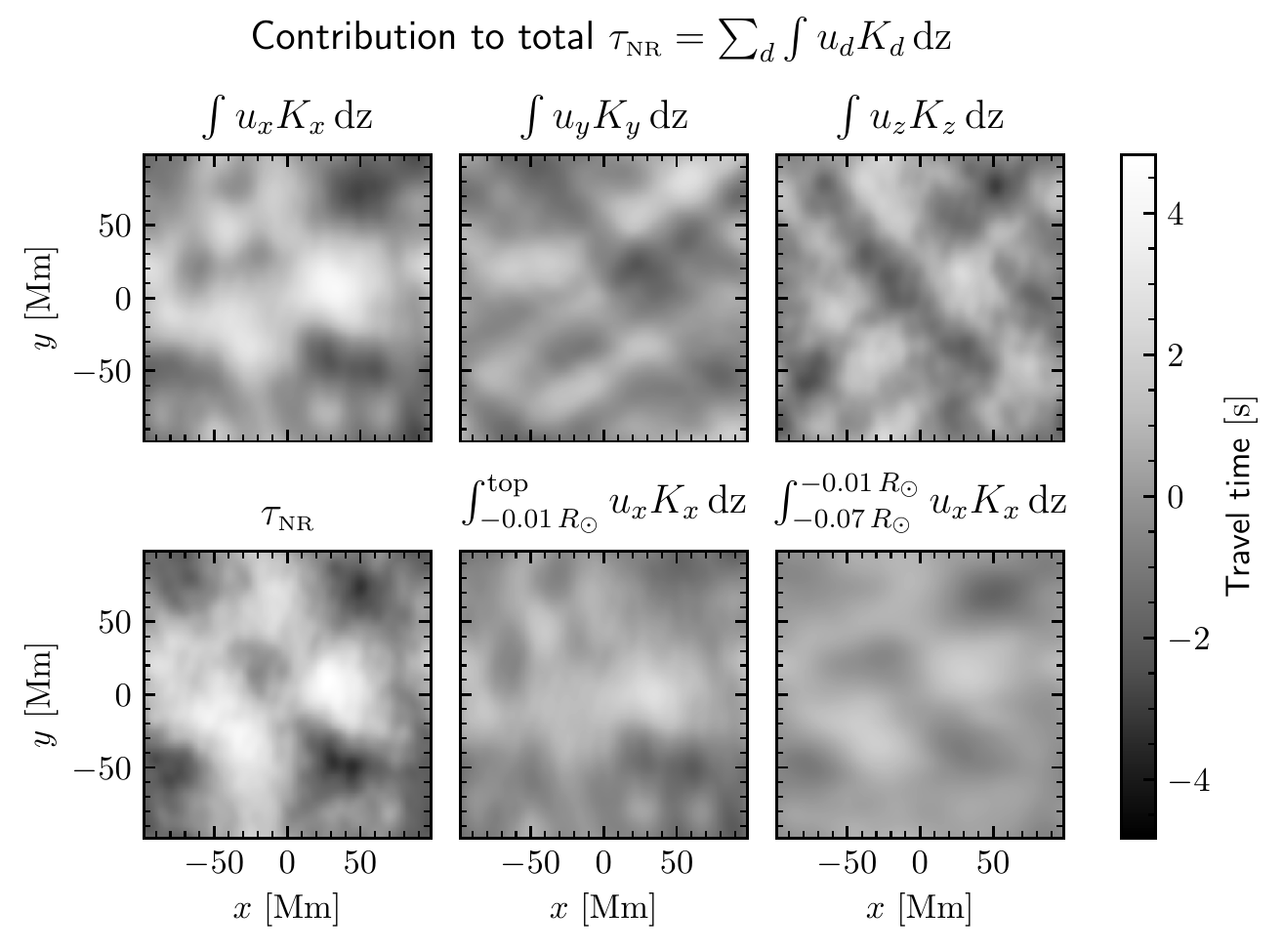}
        
        \caption{{Example noiseless travel times$\tauNR$ ({bottom} left panel) corresponding to the right panel in Figure~\ref{figNoise} and contributions from {the different flow components (top panels) as well as different} depth ranges ({remaining} panels) obtained using Eq.~\eqref{eqtauKepsCart} with $\epsilon=0$.
        }}
        \label{figExampleTau}
\end{figure}

Using these kernels and the flow fields from MURaM, we obtained forward-modeled travel-time maps using Eq.~\eqref{eqtauKepsCart}. {These travel times are free of seismic noise ($\epsilon=0$) and are thus referred to as noiseless travel times. 
{The bottom left panel in} Figure~\ref{figExampleTau} shows {an example} of these maps for one realization of the flow field. {The top panels of Figure~\ref{figExampleTau} show that $u_x$ indeed contributes most} to the travel times, {as intended by \citetalias[][]{Hanasoge2012}. However, the $u_y$ and $u_z$ flow components also contribute significantly, and a large fraction of the $u_x$ contribution comes from near-surface flows (bottom middle panel in Fig.~\ref{figExampleTau}), which was unintended. On the other hand, the bottom right panel in the figure} shows that the {signal from the} flows in the deeper layers {is} somewhat correlated with the {total signal, which shows that the deep flows might be detected using these data}.}

\subsection{Noise model and covariance of travel times}

In Cartesian geometry {and in the quiet Sun}, it is reasonable to assume that the correlation {in the noise} of two travel-time measurements only depends on their {separation} \citep[e.g.,][]{GB2004},
\begin{align}
\langle\epsilon(\bx) \, \epsilon(\bx')\rangle = \Lambda(\bx - \bx'),
\end{align}
so that, see Appendix~\ref{appGB04FFT},
\begin{align}
\langle\epsilon(\bk) \, \epsilon(\bk')\rangle = \delta_{\bk,\bk'}  \frac{\Lambda(\bk)}{h_k^2},
\end{align}
where $\Lambda$ is the noise covariance matrix and $h_k$ is the resolution in Fourier space. Here, we assumed that averages $\langle q \rangle=\langle q(t) \rangle_t$ {for any quantity $q$} are taken over many {independent} realizations $q(t)$, which are observed for different nonoverlapping periods indexed by $t$. We assumed that enough data were averaged so that the temporal averages {approximate} their expectation values.

As a consequence of the horizontal translation invariance, the problem decouples in $\bk$ space, that is,
\begin{align}
&\langle\tau^{*}(\bk) \, \tau(\bk') \rangle = \delta_{\bk,\bk'} \Bigg(  \frac{1}{h_k^2} \Lambda(\bk) \, \nonumber \\
&  + (2\pi)^4 \sum_{d,d'=r,\theta,\phi} \iint \id z  \id z' \, K^*_{d}(\bk,z) \,   K_{d'}(\bk,z')  \langle u_{d}^{*}(\bk,z)\,  u_{d'}(\bk,z') \rangle \Bigg) \label{eqtauCorr}.
\end{align}

We compute the noise covariance matrix for the deep-focus measurement using the method of \cite{GB2004} and \cite{Fournier2014} and draw realizations of the noise as described in Appendix~\ref{appGB04FFT}.

\begin{figure*}
        \centering

        \includegraphics[width=\linewidth]{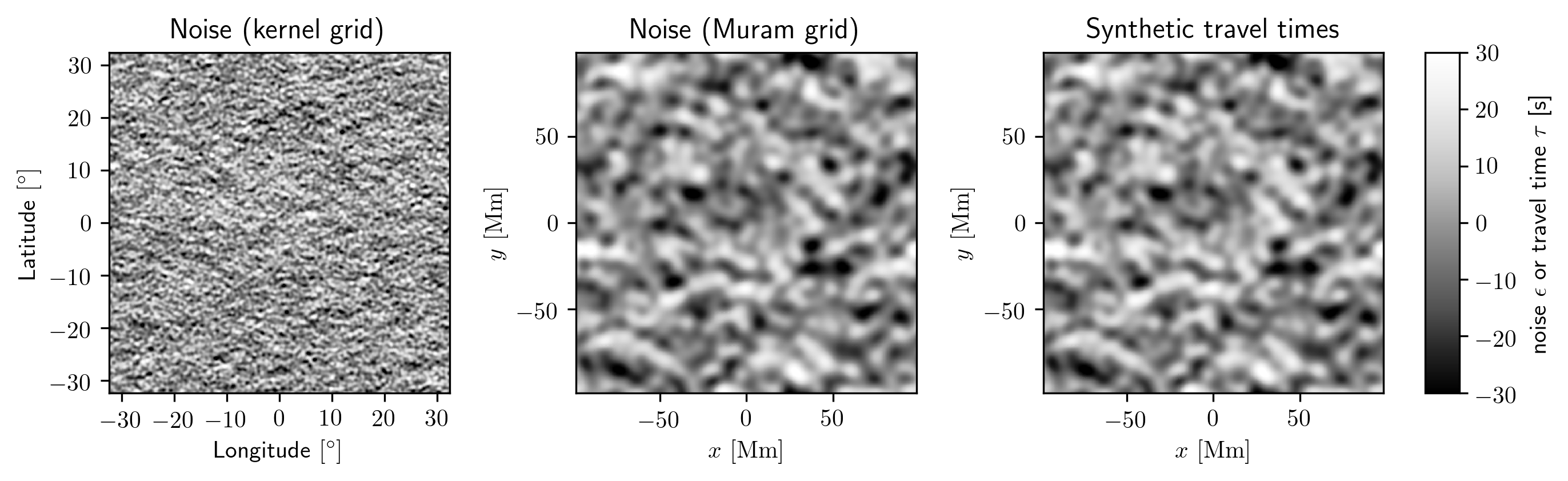}

        \caption{Example realizations of the noise $\epsilon(\bx)$ from our noise model on the kernel (left) and MURaM grids (middle), and synthetic noisy travel times $\tau(\bx)$ (right). The noise has a standard deviation of $10.3\,\rm{s}$ {and the travel-time signal in the right panel of $1.9\,\rm{s}$, which compares well with the data \citep{Hanasoge2012}.} The left panel can be compared to Fig.~5 in the supplementary material of \citet{Hanasoge2012}. {Figure~\ref{figExampleTau} shows the {contribution of the} travel-time signal {to} the right panel in more detail}.
        }
        \label{figNoise}
\end{figure*}

Figure~\ref{figNoise} shows an example realization of the {synthetic} noise. It is {qualitatively} very similar to the {actual} data{ while being a different stochastic realization}. {This can be seen by comparing the left panel in Figure~\ref{figNoise} to} Fig.~5 in the supplementary material of \citetalias{Hanasoge2012}. {Our synthetic noise has a standard deviation of $10.3\,\rm{s}$ and the contribution of the travel-time signal to the right panel of Figure~\ref{figNoise} has a standard deviation of $1.9\,\rm{s}$. Both compare reasonably well with the data \citepalias{Hanasoge2012}, where the noise has a standard deviation of $11.8\pm0.3\,\rm{s}$ and the data $2.5\pm0.02\,\rm{s}$ (\citeauthor{BirchRevisitingInPrep}, in prep.). We note that this noise model has been tested successfully against data (e.g., Figs. 5, 6, 8, 10, and 11 in \citealp{GB2004}; Fig. 5 in \citealp{Fournier2014}).}

{However, we also note that it is not clear why} Fig.~3 in \citetalias{Hanasoge2012} shows a very similar travel-time map{, but} with features aligned in north-south direction, as opposed to the features aligned in east-west direction in Fig.~5 in the supplementary information of \citetalias{Hanasoge2012} and {in our Fig.~\ref{figNoise}}. Except for the difference in the orientation, the three figures are qualitatively very similar. Because the measurements were averaged (see also Fig.~\ref{figKernels}), it is expected that travel-times are more correlated in east-west direction, as is shown in our noise map{s} {and in Fig.~5 in the supplementary material of \citetalias{Hanasoge2012}}. {On the other hand,} the orientation of the noise {does not} affect the results in our study or in \citetalias{Hanasoge2012}. {Therefore} we do not address this issue further.

{
\subsection{Assumptions for estimating an upper limit}

In this subsection, we summarize the essential assumptions leading to the estimate obtained by \citetalias{Hanasoge2012} and introduce their method. We also briefly outline the motivation for these assumptions. After summarizing the assumptions that lead to this estimate, we assess their validity in the following section. We therefore do not discuss their validity in this section. We here explicitly mark the most important assumptions made by \citetalias{Hanasoge2012} {to obtain} an upper limit.

However, we stress already here that rather than an inequality sign, an {approximate equality} sign is a more adequate assumption in many cases. In the following, we therefore include alternative formulations of the assumptions as approximations for future reference in parentheses.}

\subsubsection{Noise correction}

A major step is the so-called noise reduction, which aims at estimating the noiseless (or noise-reduced) travel times $\tauNR=\tau -\epsilon${, which is the signal of interest.} In order to {estimate} the amplitude of the signal from the noisy travel-time measurements, \citetalias{Hanasoge2012} {used a relation from \citet[][Section 6]{GB2004}. Correcting an obvious error in \citet{GB2004}, \citet{Hanasoge2012} used the correct version
\begin{align}
    \sigma^2(T) &= \frac{N_0^2}{T} + S^2, \label{eqSNfit}
\end{align}
where $\sigma^2$ is the variance of the data, $T$ is the observation time,{ $N_0^2/T=N^2$ {is the variance of the noise} for an observation time $T$}, and $S^2$ {is} the contribution of the signal to the total variance. This equation is valid in the case of time-independent flows and omits a small $1/T^2$ contribution to the noise \citep[see Eqs. (14) and (B.6) in][]{Fournier2014}. From Equation~\eqref{eqSNfit}, {it follows that}
\begin{align}
    \frac{S^2}{\sigma^2} \leq \frac{S^2}{N^2},
\end{align}
and by assuming that the signal-to-noise ratio is independent of the spatial scale, \citetalias{Hanasoge2012} obtained}
\begin{align}
\tauNR^{\rm{fft}}(\ell)  \stackrel{\rm{(A1)}}{\leq}  \sqrt{\frac{S^2}{N^2 + S^2}} \,  \tau^{\rm{fft}}(\ell)  \leq \frac{S}{N} \, \tau^{\rm{fft}}(\ell). \label{eqNoiseCorr}
\end{align}
{Here, $q^{\rm{fft}}(\ell)$ indicates an azimuthally averaged spectrum that was obtained from a horizontal Fourier transform of the quantity $q$ {(e.g., $q=\tau$ or $q=u_x$)} to be compared with the spherical notation of \citetalias{Hanasoge2012}. The quantity $q^{\rm{fft}}(\ell)$ is defined as a mode amplitude per spherical harmonic degree (same unit as $q$). As usual, an appropriate summation of $q^{\rm{fft}}(\ell)^2$ can be used to  estimate the variance in real space, see Appendix~\ref{appGB04FFT} for the exact normalization.} {{Starting} from Equation~\eqref{eqNoiseCorr}, we use a {notation like ``(A1)''} to mark inequalities or equalities that correspond to the most important assumptions from \citetalias{Hanasoge2012} that we assess in this study. {These assumptions are summarized in Section~\ref{secSummaryAssumptions}.}

{The main aspect of assumption (A1) that we assess in this study is} the assumed independence of $k$ or $\ell$ of the signal-to-noise ratio, which is implicit in Eq.~\eqref{eqNoiseCorr}. \citetalias{Hanasoge2012}} fit the signal-to-noise ratio $S/N$ in Eq.~\eqref{eqSNfit} to the data. We do not study the validity of this fitting procedure here.

In order to estimate the signal from the noisy data, {\citetalias{Hanasoge2012} used} assumption (A1), {which says} that the noise-reduced travel-time mode amplitude $\tauNR^{\rm{fft}}(\ell)$ is bounded by (or can be approximated by) the signal-to-noise ratio times the observed noisy travel-time mode amplitude $\tau^{\rm{fft}}(\ell)$. {Here, $\tauNR^{\rm{fft}}(\ell)^2$} is a {spectral} decomposition of the signal variance, and similarly, $\tau^{\rm{fft}}(\ell)^2$ is a decomposition of the total variance $\sigma^2$.

For synthetic data, we know
\begin{align}
\left(\frac{S}{N}\right)^{\rm{true}}(\ell) &=\frac{\tauNR^{\rm{fft}}(\ell)}{\epsilon^{\rm{fft}}(\ell),}
\end{align}
and we can thus compare this value with the value of $S/N$ used by \citetalias{Hanasoge2012}.

\subsubsection{Calibration curve}
\label{secCalCurve}

When the amplitude of the signal has been estimated from the data or a noiseless travel-time spectrum is given for synthetic data, the aim is to estimate the amplitude of the convective motions at the depth that generates the signal. \citetalias{Hanasoge2012} proposed to do this using a so-called calibration curve. This calibration curve simply converts the noise-reduced travel-time spectrum into a velocity spectrum.

To {introduce the derivation of} the calibration curve {proposed by \citetalias{Hanasoge2012}, including} the underlying assumptions{, w}e start from Equation \eqref{eqtauCorr}. {Assuming that the statistical properties of the data are translation invariant, {we} can only consider {correlations $\langle\tau^{*}(\bk) \, \tau(\bk') \rangle$ for} $\bk=\bk'$}. {One} then obtains\begin{align}
\langle & \tauNR(\bk)^2\rangle \nonumber \\
 &= (2\pi)^4 \sum_{dd'} \iint \id z  \id z' \,  K^{*}_{d}(\bk,z) \,    K_{d'}(\bk,z') \, \langle u_{d}^{*}(\bk,z)\,  u_{d'}(\bk,z')\rangle \label{eqTauNR}  \\
& \stackrel{\rm{(A2)}}{\geq} \,  (2\pi)^4\sum_{d} \iint \id z  \id z' \,   K^{*}_{d}(\bk,z) \,    K_{d}(\bk,z') \,  \langle u_{d}^{*}(\bk,z)\,  u_{d}(\bk,z')\rangle \label{eqFlowCorrOnlySameComponent} \\
&\stackrel{\rm{(A3)}}{\geq} (2\pi)^4  \iint \id z  \id z' \,  K^{*}_{x}(\bk,z) \,    K_{x}(\bk,z') \, \langle u_{x}^{*}(\bk,z)\,  u_{x}(\bk,z')\rangle, \label{eqFlowCorrOnlyXComponent} 
\end{align}
where{, following \citetalias{Hanasoge2012}, we assumed} that the spectrum of the travel-time signal is bounded below by (or can be estimated by) the spectrum of travel-times stemming only from the $x$-component of the flow in assumption (A3), or in an intermediate step as a sum of spectra from the three components, ignoring correlations of different flow components, in assumption (A2). These assumptions are motivated by {the idea that the measurement geometry is designed to be} mostly sensitive to the $x$ direction. { $u_x$  contributes most to the signal, see the top left and bottom left panel in Figure~\ref{figExampleTau}. However, the measurements are also sensitive to the other flow components, see Figure~\ref{figKernels} and the top middle and top right panels in Figure~\ref{figExampleTau}. We evaluate and discuss the validity of the assumptions made by \citetalias{Hanasoge2012} and the effect on the inferred flow spectrum in detail in Section~\ref{secResults}.}

 {\citetalias{Hanasoge2012} assumed} that the flow near the target depth of the travel-time measurement bounds the spectrum from below (contributes dominantly to the spectrum of travel times). In addition, {they} assumed that the correlation of the flow at different depths $\langle u_{x}^{*}(\bk,z)\,  u_{x}(\bk,z')\rangle$ may be bounded (or modeled) by a Gaussian, more specifically that some Gaussian is centered at {the target depth} $z_T = (0.96-1)R_\odot$ ($z=0$ is at $r=R_\odot$) with a width of $\sigma_z'=16.77\,\rm{Mm}=1.8 H_P${, where $1.8$ pressure scale heights $H_P$ is the mixing length. It is required here that the Gaussian} falls off faster than the correlation function, such that
\begin{align}
\langle u_{x}^{*}(\bk,z)\,&  u_{x}(\bk,z')\rangle \geq \nonumber \\
 &\langle |u_x(\bk,z_T)|^2\rangle \,\exp(-\frac{(z-z_T)^2}{2\sigma_z'^2}) \, \exp(-\frac{(z'-z_T)^2}{2\sigma_z'^2}), \label{eqVxCorrelationLength}
\end{align}
and consequently, if $K^{*}_{x}(\bk,z) \,    K_{x}(\bk,z') >0$,
\begin{align}
\text{Eq. \eqref{eqFlowCorrOnlyXComponent}}
&\stackrel{\rm{(A4)}}{\geq} |\cK(\bk,\sigma_z')|^2 \, \langle |u_x(\bk,z_T)|^2\rangle,\label{eqAssumptionAroundTarget}
\end{align}
where
\begin{align}
\cK(\bk,\sigma_z') &= (2\pi)^2 \int \, \id z  \,   K_{x}(\bk,z)   \,\exp(-\frac{(z-z_T)^2}{2\sigma_z'^2}). \label{eqKtilde}
\end{align}
{We did not confirm whether the assumption of a Gaussian shape of the correlation function is an appropriate choice. This is left to future studies.}

The final assumption is
\begin{align}
        \frac{1}{N_\ell} \sum_{\bk \in I_\ell}   \langle |u_x(\bk,z_T)|^2\rangle \, |\cK(\bk,\sigma_z')|^2 &\stackrel{(A5)}{\geq} 
        \left(\frac{1}{N_\ell} \sum_{\bk \in I_\ell}   \langle |u_x(\bk,z_T)|^2\rangle \right) \, C_\ell'^2, \label{eqAzimuthalK}
\end{align}
where $I_\ell=\{\bk \text{ such that } |\bk|-h_k/2 \leq \ell/r < |\bk|+h_k/2 \}$ selects all {horizontal wave vectors} of similar length, $N_\ell$ {is} the number of elements in $I_\ell$,
\begin{align}
C_\ell'^2  &= \frac{1}{N_\ell} \sum_{\bk \in I_\ell}    |\cK(\bk,\sigma_z')|^2, \label{eqCl}
\end{align}
and $\ell=\ell_j$ takes the values $\ell_j=j\,h_k r$ with integer $j$ in order to be compatible with the discrete Fourier grid. This introduces the so-called calibration curve, $C_\ell'$. {By making assumption (A5) in Eq.~\eqref{eqAzimuthalK}, {we assume} that the azimuthal dependence of the flow and the kernel spectra are not anticorrelated, such that the azimuthal average of the kernels and flow spectra can be taken independently, without increasing the result. This was assumed but not explicitly mentioned by \citet{Hanasoge2012}.}

By taking the azimuthal average of the travel-time power spectrum in Eq.~\eqref{eqTauNR} and by using the above assumptions, {one obtains} the final inequality on which the estimate of the upper limit of the flow amplitude by \citetalias{Hanasoge2012} is based. {At the same time, we convert from Cartesian into spherical geometry in the Fourier domain using $\ell=k r$ and $h_\ell = h_k r$. Here and in the definition of $I_\ell$ above, we used $r\approx \rm{const} = R_\sun$ for the entire domain, which is equivalent to the assumption of Cartesian geometry. In addition, we employed} the convention for converting spectra from Fourier space into spherical harmonic space outlined in Appendix~\ref{appGB04FFT}, which is inspired by \citeauthor{BirchRevisitingInPrep} (in prep.),
\begin{align}
\tauNR^{\rm{fft}}(\ell)^2 &= \frac{h_k^3}{(2\ell+1)r}\sum_{\bk \in I_\ell}  |\tauNR(\bk)|^2 \\
&\stackrel{\text{Eqs.~\eqref{eqFlowCorrOnlyXComponent},\eqref{eqAssumptionAroundTarget},\eqref{eqAzimuthalK}}}{\geq}  \frac{ h_k^3}{(2\ell+1)r}\sum_{\bk \in I_\ell}   \langle |u_x(\bk,z_T)|^2\rangle \,  \, C_\ell'^2\label{eqbla} \\
&= u^{\rm{fft}}_x(\ell,z_T)^2   \, C_\ell'^2. \label{eqtauCl}
\end{align}

In order to facilitate the numerical computation of Eq.~\eqref{eqKtilde}, \citetalias{Hanasoge2012} instead chose a smaller width $\sigma_z= \sigma_z'/D$, where $D=9.64$, see also \citeauthor{BirchRevisitingInPrep} (in prep.). {As \citetalias{Hanasoge2012} derived $\cK$ or $C_\ell'$ from numerical simulations of wave propagation, shrinking the correlation width by the factor $D$ permitted them to deduct a larger number of independent measurements from one simulation, and thus to reduce the numerical cost. However, the quantity $D$ at the same time acts as an additional free parameter. We show below that its introduction has important effects on the estimated spectrum.}

After obtaining $\cK(\bk,\sigma_z)$ and the alternative calibration curve
\begin{align}
C_\ell &= \sqrt{\frac{1}{N_\ell} \sum_{\bk \in I_\ell}    |\cK(\bk,\sigma_z)|^2}, \label{eqClHan12}
\end{align}
\citetalias{Hanasoge2012} then approximated $C_\ell'$ by
\begin{align}
C_\ell' \stackrel{\rm{(A6)}}{=} D\, C_\ell. \label{eqClprimeDCl}
\end{align}

The integration around the target depth with a Gaussian weighting of width $\sigma_z'$ is equivalent to assuming that the flows are vertically correlated with each other for about a scale height. This is reasonable for the simulated data \citep{Lord2014}, although a scale dependence of this width may be a more adequate assumption. At the same time, the value of the vertical correlation length of the flows is not known for the Sun.

\subsubsection{Summary of assumptions}
\label{secSummaryAssumptions}

As a result of the above assumptions, we obtain, see Eqs.~\eqref{eqNoiseCorr},~\eqref{eqtauCl}, and~\eqref{eqClprimeDCl},
\begin{align}
u^{\rm{fft}}_x(\ell,z_T) \stackrel{(A2)-{(A6)}}{\leq}  \frac{\tauNR^{\rm{fft}}(\ell)}{D\,C_\ell} \stackrel{(A1)}{\leq}  \frac{S}{N} \, \frac{\tau^{\rm{fft}}(\ell)}{D\,C_\ell} = \hat u^{\rm{fft}}_x(\ell,z_T), \label{eqUest}
\end{align}
{where the equivalent of the right-hand side was used by \citetalias{Hanasoge2012} to estimate the flow near the target depth. We write $\hat u_x$ for the estimate of $u_x$.} We now briefly summarize the most important assumptions that were explicitly marked above and that were proposed by \citetalias{Hanasoge2012}{. We evaluate them in the following section}.
\begin{itemize}
        \item[(A1)] {T}he variance in the travel-time maps stemming from the signal and due to the flow is bounded above by (or can be estimated by) obtaining the S/N from a fit to the data \cite[see][in prep.]{Hanasoge2012}. {The S/N} is independent of spatial scale, see Eq.~\eqref{eqNoiseCorr}.
        \item[(A2)] {T}he noise-reduced spectrum of travel times is bounded from below by (or can be estimated by) taking only correlations of a flow component with itself into account, see Eq.~\eqref{eqFlowCorrOnlySameComponent}.
        \item[(A3)] {T}he spectrum of travel times is further bounded from below by (or can be estimated by) taking only correlations of the $x$-component of the flow with itself into account, see Eq.~\eqref{eqFlowCorrOnlyXComponent}. This assumption stems from the observation that the travel-time geometry is chosen such that the major contribution is in fact from this component.
        \item[(A4)] {T}he spectrum of travel times is further bounded from below by (or can be estimated by) the contribution of the dominant depth range, which is assumed to be a pressure scale height around the target depth of the deep-focus measurement, see Eq.~\eqref{eqAssumptionAroundTarget}.
        \item[(A5)] {T}he azimuthal dependence of $u_x$ and $K_x$ is independent, that is, it does not matter whether these quantities are first multiplied and are then averaged over the azimuth of \textbf{\textit{k}}$\bk$, or whether the average is taken first and is then multiplied, see Eq.~\eqref{eqAzimuthalK}.
        \item[(A6)] To obtain the calibration curve, similar results are obtained regardless of whether the kernels are integrated over about a pressure scale height, $\sigma_z'$ or over  the smaller width $\sigma_z=\sigma_z'/D$ and then the result is multiplied by $D$, that is, $C_\ell' \approx D\, C_\ell$, see Eq.~\eqref{eqClprimeDCl}.
\end{itemize}
Furthermore, {a few more} assumptions {were (at least implicitly) made by \citetalias{Hanasoge2012}. W}e state {these assumptions here }for completeness; {these assumptions are not be studied here but should be addressed in future work}:
\begin{itemize}
        \item[(B1)] In fitting the signal-to-noise ratio, \citetalias{Hanasoge2012} {followed the assumption of \citet{GB2004} that} the noise decreases with $1/\sqrt{T}$. A potential $1/T$ term \citep[see Eqs. (13) and (B.6) in][]{Fournier2014} is thus neglected.
        \item[(B2)] The effect of a time dependence of the convective flows, either on the fit of the signal-to-noise ratio through a time dependence of the signal, or on the travel times beyond a Born-approximation model for constant flows \citep{GB2002}, can be neglected. {Because convective flows are time-dependent, time-dependent perturbation theory should be used in an ideal scenario.}
        \item[(B3)] The convolution theorem can be used. This is straightforward for Cartesian geometry, see Eq.~\eqref{eqtauKepsCart}, but might be inaccurate for spherical geometry.
        {\item[(B4)] The vertical correlation of the flow components can be approximated (or bounded) by a Gaussian, see Eq.~\eqref{eqVxCorrelationLength}. The possibility of an anticorrelation of divergent flow components (due to mass conservation) is thus not taken into account, but should be modeled in future work.}
\end{itemize}

\section{Results}
\label{secResults}

\subsection{Scale-dependent noise correction is important}

\begin{figure}
        \centering

        \includegraphics[width=\linewidth]{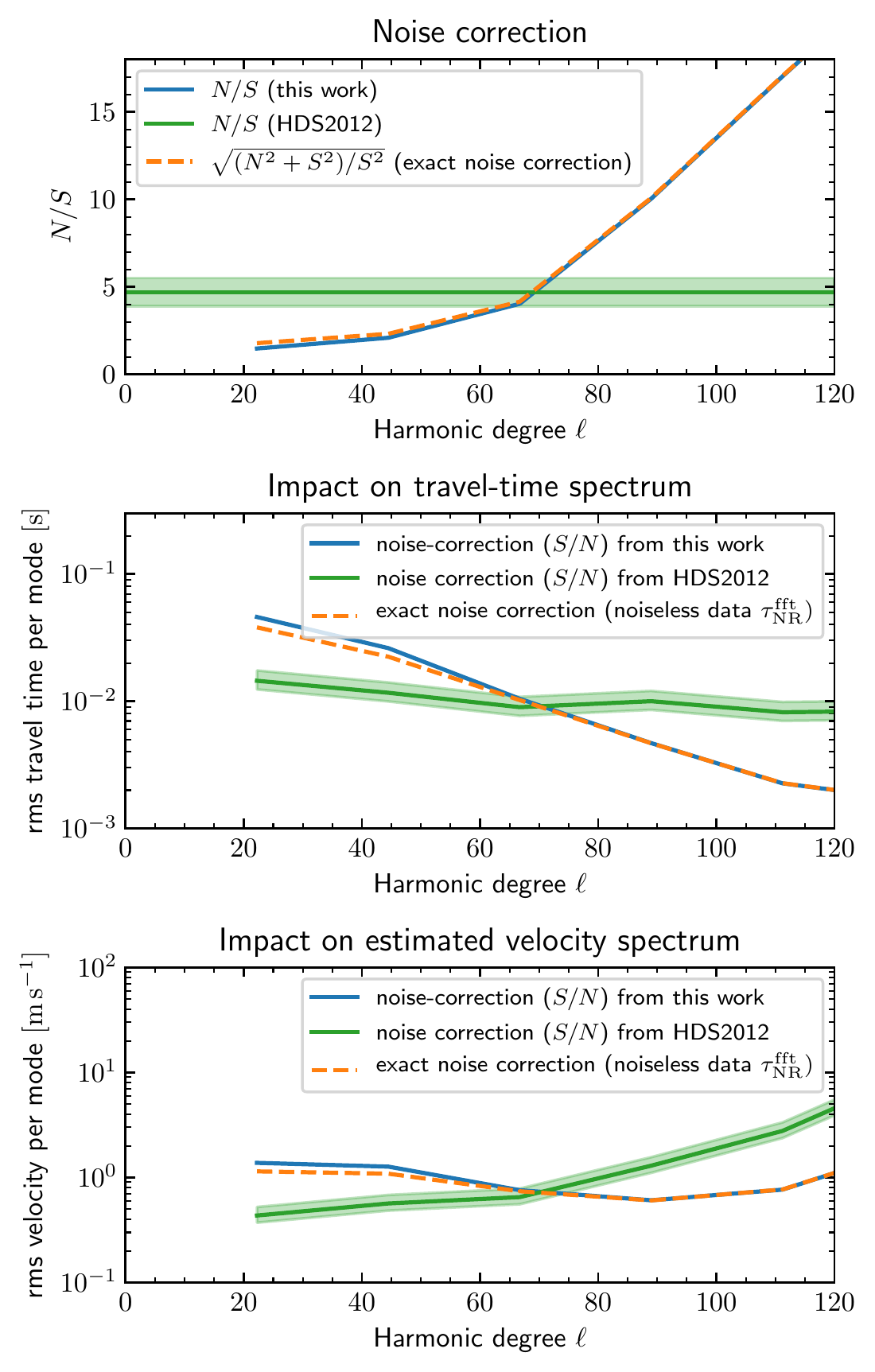}

        \caption{{Inverse of} the signal-to-noise ratio (top) and its effect on the noise reduction (middle) as well as the estimate of the spectrum of horizontal flows (bottom, as a velocity amplitude per mode) for different approaches. The signal-to-noise ratio is clearly scale dependent, in contradiction to assumption (A1). {We use $\ell=k R_\sun$ to convert from wave number into harmonic degree. We do not show results for $\ell=0$ because this corresponds to an arbitrary change in reference frame.}}
        \label{figNoiseCorr}
\end{figure}

For MURaM-type flows, we find that the signal-to-noise ratio significantly depends on the spatial scale and is not independent of scale, as assumed by \citetalias{Hanasoge2012}, see the top panel in Figure~\ref{figNoiseCorr} and assumption (A1). The true signal-to-noise ratio for the Sun may be different to what we report here if the subsurface flows in the Sun are substantially different from MURaM flows. We strongly recommend that the noise correction be made as a function of harmonic degree (or wave number){, possibly even as a function of azimuth (i.e., $m$ or $k_x/|\bk|$).}

The middle and bottom panels in Figure~\ref{figNoiseCorr} show the effect of this scale-dependent noise correction on the spectrum of travel times and the estimated flow amplitudes. For MURaM flows, a constant signal-to-noise ratio for all scales clearly leads to an underestimation of the flow amplitudes at the largest scales.

{\subsection{Travel times are sensitive to near-surface flows and to all flow directions}
\label{secContribSignal}

The deep-focus travel-time geometry of \citetalias{Hanasoge2012} is designed to be sensitive mostly to $u_x$ flows (the east-west component of the flow) around the target depth of $0.96\,R_\sun$. Figure~\ref{figKernels} shows that the travel times are in principle sensitive to all flow components and to a broad range of depths, including flows near the surface. However, there is some focus on the target depth for the sensitivity to $u_x$ flows, see Figure~\ref{figKernels} (bottom left panel).

This is also reflected in the contribution of the different flow components and depth layers to the travel-time signal, which we show in Figure~\ref{figExampleTau}. Although $u_x$ (top left) contributes most to the signal (bottom left), $u_y$ and $u_z$ also contribute significantly (top center and right). The contribution from the surface layers (bottom center) is also large. However, the signal from the deeper $u_x$ flows (bottom right) is visible in the signal, which means that $u_x$ flows around the target depth might be detected using this method.}

\subsection{Calibration curve can be reproduced, but depends on assumed flow properties}

Second, we find that the computation of the calibration curve, which was done by \citetalias{Hanasoge2012} using simulations of wave propagation, can be reproduced, see the blue line in Figure~\ref{figCalCurve}. The agreement is quite good when compared to Fig.~2 of the supplementary material of \citetalias{Hanasoge2012}, even when we computed it on the coarser-grained {wave number} grid of the MURaM simulation, see the green dashed line in Figure~\ref{figCalCurveDetail}.

However, \citetalias{Hanasoge2012} computed the calibration curve assuming a width of convective features of $\sigma_z = 1.74\,\rm{Mm}$ and later rescaled the result by a factor of  $D=9.64$ to mimic an effective width of $\sigma_z'=D\sigma_z${, which is similar to the mixing length or $1.8$ pressure scale heights} at $0.96R_\sun$ (see {Sec.~\ref{secCalCurve} and} \citeauthor{BirchRevisitingInPrep}, in prep.). We compare this to directly assuming a larger width of convective features of $\sigma_z'=D\sigma_z$, see $C_\ell'$ in Figure~\ref{figCalCurve}. This alternative calibration curve is lower. Using $C_\ell'$ thus results in a correspondingly higher estimated upper limit, see Eq.~\eqref{eqUest}.

In other words, the use of the width $\sigma_z=\sigma_z'/D$ and rescaling the calibration curve by using $D C_\ell$ as done by \citetalias{Hanasoge2012} effectively results in a lower estimate for the upper limit. This difference arises because the kernel is not constant in the area $\sigma_z'^2$ around the target depth, see Fig.~\ref{figKernelkzzprime}. This increases the value of $D C_\ell$ compared to $C_\ell'$. Consequently, assumption (A6) is thus not met and the resulting inequality is in the wrong sense. As the value of $\sigma_z$ is an assumption on the vertical correlation length of the flow, an unknown property of the Sun enters the method here. {In addition, the vertical correlation function is approximated by a Gaussian, although its shape is not known, see the discussion in Section~\ref{secVertCorrStruct}.}

The introduction of the rescaling with $D$ acts similar to a free parameter. The overall effect of this is further discussed in section~\ref{secWhyWorks}.

\begin{figure}
        \centering
        \includegraphics[width=\linewidth]{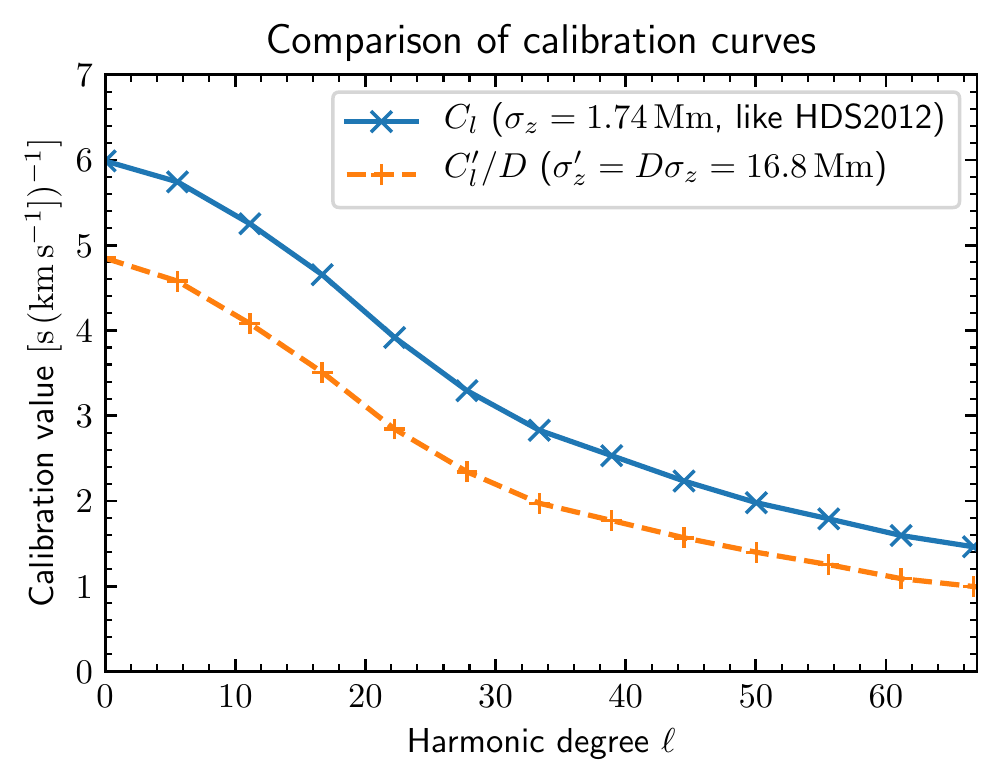}
        \caption{Calibration curve of \citet[][Fig.~2 in the supplementary material]{Hanasoge2012} can be reproduced (blue solid curve). When a more realistic width $\sigma_z'=16.8\,\rm{Mm}=1.8\,H_P$ of flow features is used to compute the calibration curve, a change is seen (dashed orange curve), which would lead to an increase in the estimated upper limit, see assumption (A6). {The crosses and pluses indicate the grid points in harmonic degree, which were obtained by converting horizontal Fourier modes from Cartesian to spherical geometry (see Appendix~\ref{appSHT}). The kernels and thus the calibration curves shown here were computed on a spatial grid with a four times larger spatial extent than the MURaM grid, therefore the resolution in Fourier space, $h_l=r h_k = r \frac{2\pi}{L}$, is four times finer than in other plots in this paper. See Fig.~\ref{figCalCurveDetail} for a more detailed comparison.}}
        \label{figCalCurve}
\end{figure}

\subsection{Concept of a calibration curve is generally valid}
\label{secResultsCalCurve}

Furthermore, we confirmed the assumptions (A2)-(A6) that were made to derive the calibration curve, see Figure~\ref{figCheckCalCurve3plots}. While some of the inequalities were met, some were violated, although not substantially. We conclude from this that the concept of a calibration curve is generally valid when the resulting estimate is rather understood as a rough estimate of the amplitude of the flow and not a strict upper limit. {We recall that the calibration curve depends on the assumed vertical correlation function, which is not known for the Sun, see the discussion in Section~\ref{secVertCorrStruct}.}

\subsubsection{Azimuthal averages have to be treated more carefully}
\label{secAzimuthalAvg}

We find that when the effect of off-diagonal flow-correlation is neglected, assumption (A2) is not justified because the off-diagonal elements make a significant negative contribution, see Figure~\ref{figKKuu066}. The reason for this is as follows. When the dependence of $K_x^* K_y$ and $\langle u_x^* u_y \rangle$ on the direction of \textbf{\textit{k}}$\bk$ is considered, we observe that the two quantities are anticorrelated, see {the left panel in} Figure~\ref{figKKuuazim066}. This is the case for almost all $k$ and results in a negative total effect of the off-diagonal terms on the travel-time spectrum. It might be possible to remove this anticorrelation in future studies by using a {pure} point-to-point {travel-time geometry, where cross correlations are not averaged over arcs as done {by} \citetalias{Hanasoge2012}}, or by considering a decomposition in divergent and vortical flows, rather than $x$ and $y$ components{, see also the discussion in Section~\ref{secConlusions}}.

\begin{figure*}
        \centering
        \includegraphics[width=1\linewidth]{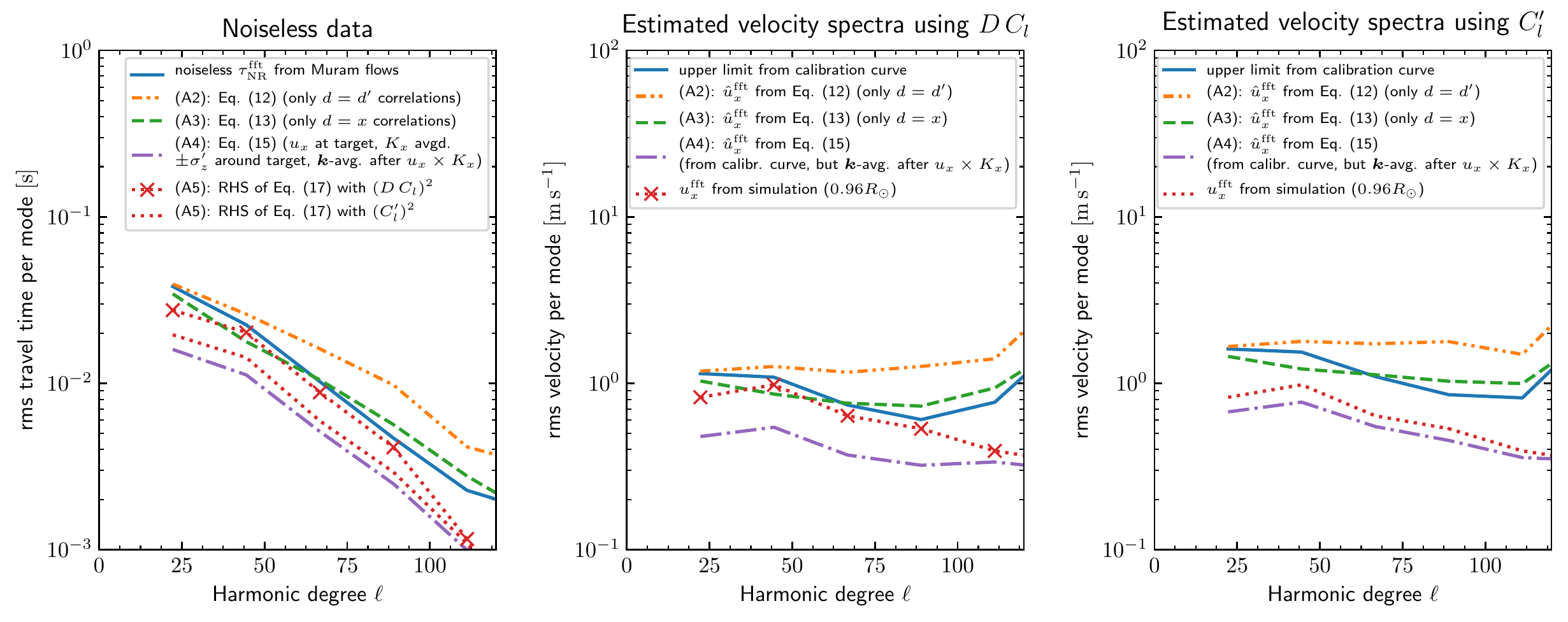}
        \caption{Test of the validity of the assumptions that {enter the derivation} of the calibration curve, shown as a noise-reduced signal in travel times (left panel) and flow amplitude (center and right panels) as a function of scale. If all assumptions (A2)-(A5) were satisfied, the {different} curves {in each panel} would {not cross each other, but be ordered from higher to lower values} in the same sense as {they are ordered from top to bottom} in the legends (e.g., $\tauNR^{\rm{fft}}\geq \text{Eq. \eqref{eqFlowCorrOnlySameComponent}} \geq \text{Eq. \eqref{eqFlowCorrOnlyXComponent}}\geq \text{Eq. \eqref{eqAssumptionAroundTarget}} \geq \text{Eq. \eqref{eqAzimuthalK}}$ for the left panel). }
\label{figCheckCalCurve3plots}
\end{figure*}

We find that the diagonal elements have a positive contribution (everywhere but for a small region for $d=z$, see Figure~\ref{figKKuu066}). As a consequence, neglecting $d=d'=y$ and $d=d'=z$, as in assumption (A3), is justified {to obtain an upper limit. However, we find that all flow components contribute significantly to the travel-time spectrum, see Figure~\ref{figCheckCalCurve3plots} (compare the blue, orange, and green curves in the left panel). This is consistent with all flow components contributing to the travel-time signal (see Sec.~\ref{secContribSignal}). To obtain an accurate estimate of the flow spectrum, the effect of all flow components thus has to be taken into account.}

Similar to the off-diagonal case, neglecting the dependence of $K_x^* K_x$ and $\langle u_x^* u_x \rangle$ on the direction of $\bk$ introduces an error, see {the right panel in} Figure~\ref{figKKuuazim066}. The magnitude of the error depends on $k$ and $r/R_\sun$, but it is generally about 20\% (as the effect on the spectrum, not on the mode amplitude). The resulting effect on the travel-time spectrum is of the wrong sign and artificially increases it, see Figure~\ref{figCheckCalCurve3plots}. Strictly speaking, assumption (A5) is thus not justified.

\subsubsection{Major contribution from {near-}surface flows, small contribution from target depth}
\label{secMostlySurface}

We find that the major contribution to the travel-time spectrum comes from {near} the surface. Only about 10-30\%, rarely up to 40\%, of the travel-time spectrum is produced below $0.99R_\sun$ or in a Gaussian envelope with width $\sigma_z'$ around the target depth, see Figure~\ref{figcontribtauk2below099rsun}. Assumption (A4) is thus well satisfied as an inequality, but is far from being a good approximation. This result is consistent with the {sensitivity kernels peaking} near the surface {(Fig.~\ref{figKernels}) and with the travel-time signal having a dominant contribution from the surface (Sec.~\ref{secContribSignal}).}

\subsection{Rather a rough estimate than a strict upper limit}

\begin{figure}
        \centering
        \includegraphics[width=\linewidth]{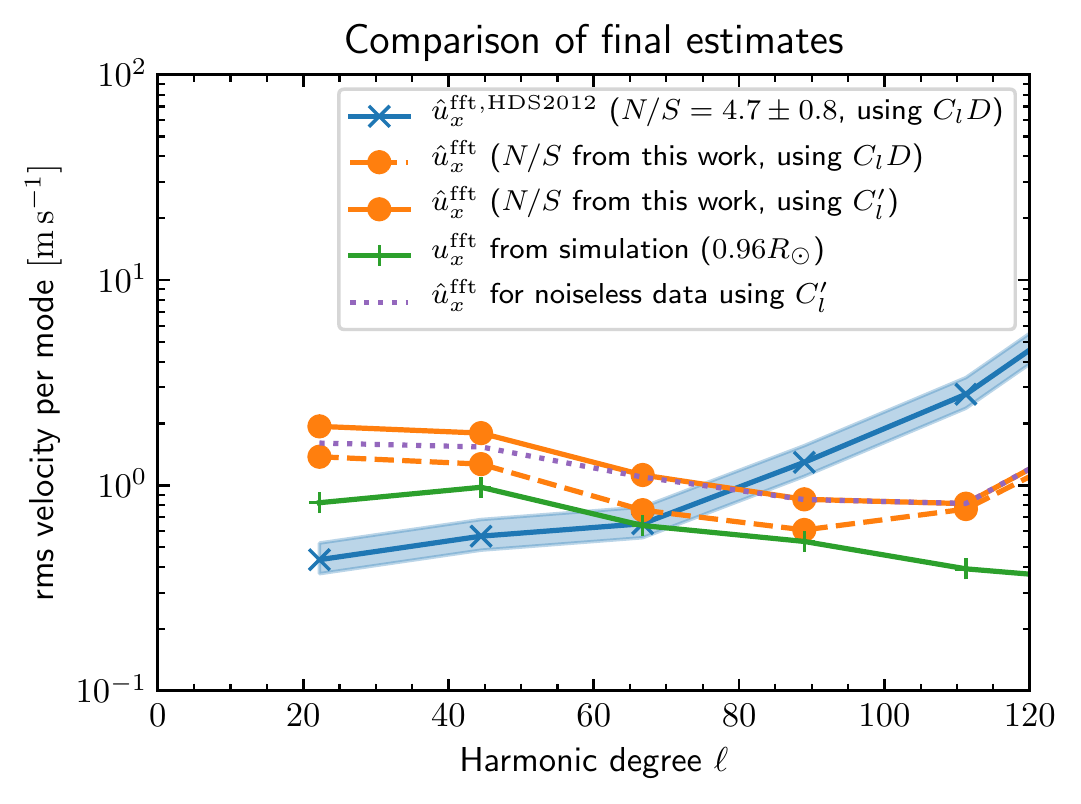}
        \caption{Comparison of estimates of the spectrum of convective flows obtained from the synthetic data based on flows from a MURaM simulation. The method of \citetalias{Hanasoge2012} applied to this synthetic data (solid blue line) can be improved by using a scale-dependent signal-to-noise correction (dashed orange line, this work). Both estimates give the correct order of magnitude of the flow (green line), although the estimate from the method of \citetalias{Hanasoge2012} is not a strict upper limit. When the underlying assumptions are improved (solid orange line), our best estimate results in a larger upper limit.}
        \label{figUest}
\end{figure}

Finally, we compared the resulting estimates of the flow amplitude with the actual estimates in Figure~\ref{figUest}. Taking all previous results into account, we find that it is more adequate to consider the estimates to be rough estimates of the magnitude of the flow than a strict upper limit. Equivalently, we find it to be more accurate to write an approximat{e equality} sign instead of an inequality in most cases for assumptions (A1)-(A6).

Figure~\ref{figUest} compares the estimates
\begin{align}
\hat u^{\rm{fft},\rm{HDS2012}}_x(\ell,z_T) &=  \left( \frac{S}{N} \right)^{\rm{HDS2012}} \frac{\tau^{\rm{fft}}(\ell)}{D \, C_\ell}, \\
\hat u^{\rm{fft}}_x(\ell,z_T) &= \left(\frac{S}{N}\right)^{\rm{true}} \frac{\tau^{\rm{fft}}(\ell)}{D\,C_\ell},
\end{align}
with the amplitude of the time-averaged flow, $u^{\rm{fft}}_x(\ell,z_T)${, which was averaged over six snapshots that were taken within the period of one day (see Sec.~\ref{secMURaM})}. {We} averaged all estimates and actual flow amplitudes over the 11 available realizations. These averages were taken over the corresponding spectra before the square root was taken to obtain the flow amplitude.

\subsection{Why the method gives the correct order of magnitude as a result}

\label{secWhyWorks}

Considering the results shown in Figure~\ref{figUest}, we conclude that for MURaM-like flows, the method of \citetalias{Hanasoge2012} just as applied in that paper (blue curve) gives the correct order of magnitude of the flow within a factor of about two {for the larger scales with $\ell\lesssim 90$. For scales with $\ell<60$, the estimate is not an upper limit because it underestimates the true flow by a factor of about two. For the smaller scales with $\ell>100$, the estimate is too large by nearly an order of magnitude, although it is a true upper limit}.

When the noise correction is applied correctly {and the free parameter $D$ is eliminated} (solid orange curve), the agreement is better{. In nearly the entire range ($\ell<110$), the improved estimate} constitutes an upper limit{ and is consistent with the actual flow spectrum within a factor of two}.

Given that {for both cases} some of the inequalities are not met, see Figure~\ref{figCheckCalCurve3plots} and Section~\ref{secAzimuthalAvg}, the question remains why the method {works, at least for estimating the order of magnitude}. To understand this, we first note that considering only the effect of $d=x$ flow correlations roughly gives the right amplitude of the travel-time spectrum. This is because the cumulative effects of the other flow components are partly positive and partly negative, and roughly cancel out, see {Figure~\ref{figCheckCalCurve3plots} (blue versus green curve in left panel) and} Figure~\ref{figKKuu066} (cumulative effect of assumptions A2 and A3).

When we only take the flow near the target depth into account, the travel-time spectrum is greatly reduced, see Figure~\ref{figcontribtauk2below099rsun}. The flows around the target depth only contribute about 10-40\% to the squared signal. See also the (A4) curve in Figure~\ref{figCheckCalCurve3plots} and the corresponding estimate in the right panel in Figure~\ref{figCheckCalCurve3plots}. This effect is almost completely compensated for by the amplification from the change from the calibration curve $C_\ell'$ to $D C_\ell$, see middle panel in Figure~\ref{figCheckCalCurve3plots}. By this change, a free parameter $D$ is effectively introduced {by \citetalias{Hanasoge2012}}, which compensates for the inaccuracies that occurred previously. We can only speculate about the reason why th{e} compensation {using $D$} is apparently of the correct magnitude{ to yield a tight upper limit, as can be seen in the middle panel of Figure~\ref{figCheckCalCurve3plots} and for the dashed line in Fig.~\ref{figUest}}. It may be due to chance or arise because \cite{Hanasoge2010} and \citetalias{Hanasoge2012} calibrated the inferred spectrum of convection with the flows from simulations, and hence intuitively chose a value for $D$ that gives the correct magnitude of the estimated spectrum for the flow {field used in their study}.

{For smaller scales ($\ell>110$), the estimates become worse, as can be seen toward the right end of Figures~\ref{figUest} and~\ref{figCheckCalCurve3plots}. This is because for increasingly smaller horizontal scales, the sensitivity of the travel times to flows near the target depth decreases and the sensitivity to near-surface flows increases.}

{\subsection{Importance of the vertical correlation structure and the surface contribution}

\label{secVertCorrStruct}

Because the signal receives a major contribution from near-surface flows (Sections~\ref{secContribSignal} and~\ref{secMostlySurface}, Fig.~\ref{figcontribtauk2below099rsun}), a correct conversion of a travel-time spectrum to a flow spectrum always has to take the relation of the surface flows to the deep flows into account, see especially Figure~\ref{figKKuu066}. It would be possible to measure this correlation function for MURaM flows, but the true vertical correlation function for the Sun is not known. We therefore find as a major conclusion that the vertical correlation structure of the flow is needed as a prior knowledge for this method to be usable, or equivalently, that its result crucially depends on the assumed correlation structure.

This point concerns not only the correlation of the surface with the deeper regions, but also the relative amplitude of the deep flows compared to those closer to the surface. In other words, without knowing the effect of the near-surface layers to the travel-time spectrum, it is impossible to infer the spectrum at greater depths. Inferences of the spectrum at depth from observations can thus only be trustworthy when the spectrum is inferred at several depths at once, for example, using information from different target depths and an inversion procedure. Using different target depths would in principle also include information on the vertical correlation structure.

Alternatively, it may be possible to improve the observational setup so that the sensitivity kernel is better focused at the target depth (see the bottom left panel in Fig.~\ref{figKernels} for the current lack of focus and Sec.~\ref{secContribSignal}). If the measurement were only sensitive to flows close to the target depth, the problem discussed above would be avoided. Reducing the sensitivity to near-surface flows would be a first step toward such an improvement.

}

\section{Conclusions}
\label{secConlusions}

We have performed an independent evaluation of the method suggested by \citetalias{Hanasoge2012} to estimate an upper limit of solar convective flow velocities from helioseismic measurements at a target depth of $r=0.96 R_\sun$. We did this using synthetic helioseismic data for which the true flow is given by the output of a larger scale MURaM MHD convection simulation {of the quiet Sun \citep[][]{Lord2014thesis}} similar to \cite{Lord2014}. Furthermore, we used consistent normalizations of the relevant spherical harmonic and Fourier transforms as proposed by \citeauthor{BirchRevisitingInPrep} (in prep.).

For the case of MURaM flows, we find that the estimates obtained using this method give the correct order of magnitude of the actual flow amplitudes within a factor of two. However, at the largest scales, the actual flow amplitude is underestimated and the upper limit is too low by a factor of about two. This is due to the scale dependence of the signal-to-noise ratio, which was not considered in the original study \citepalias{Hanasoge2012}.

Because several simplifying assumptions enter the method we studied, the question arises why it roughly works. One way to look at this is by considering individual effects of every assumption made. When the correction for the signal-to-noise ratio is made, we find that when only the $x$ component of the flow is considered, this is a reasonable approximate model for the travel-time spectrum{, although the other components contribute significantly as well}. Even though the flows around the target depth then only contribute 10-40\% to the travel-time spectrum, the effect of this on the estimated upper limit is nearly compensated for by a rescaling of the calibration curve introduced by \citetalias{Hanasoge2012}. As a result, the estimated upper limit is just above the actual flow spectrum. If the rescaling of the calibration curve were not done, the upper limit would be higher by a factor of about 1.5. The dilemma is that the answer to how the calibration curve is correctly computed depends on the vertical correlation length of the actual flows in the Sun{, the shape of the vertical correlation function,} and the ratio of the amplitudes at depth to the surface, none of which are known. Another way to look at this is as follows. For MURaM flows, even though some inequalities that each corresponds to an assumption are not met, others are met with substantial {leeway}, so that the final estimate still is an upper bound, except at large scales because of the scale dependence of the signal-to-noise correction.

Several lessons can be learned from our study that should be taken into account in future inferences of the flow spectrum at depth. The first is the mentioned scale dependence of the signal-to-noise ratio.
        
Second, the uncertainty of how the calibration curve is to be computed is substantial. This is because the correct calibration curve depends on the properties of the actual flows in the Sun, {especially the vertical correlation function of the horizontal flows,} which {is} not known. {In addition, the travel-time measurements are more sensitive to near-surface flows than to flows near the target depth. As the ratio between the near-surface spectrum and the spectrum at depth is not known, the effect of the near-surface flows on the measured travel-time spectrum is hard to infer from observations of only one target depth.} Possible solutions to this may be performing inversions using several target depths to infer the vertical correlation structure to improve the goodness of the focus at depth (e.g., using helioseismic holography, \citealp{Lindsey2000}), or to fit several different models for power spectra and vertical correlation lengths to the data.

Third, the dependence of the kernel and the flow on the azimuthal angle in the Fourier domain has to be taken into account. As a result of this dependence, correlations of $u_x$ with $u_y$ have non-negligible effects on the travel-time spectrum. A solution may be decomposing the flow into divergent and vortical components, rather than $x$ and $y$, or using point-to-point or point-to-annulus geometry rather than an arc-to-arc {travel-time} geometry. Alternatively, because the problem decouples in $\bk$ space, an inversion may solve for every \textbf{\textit{k}}$\bk$ individually.

This study may be extended in future work. First, the effect of the time dependence of the flow has not been studied, neither on the signal-to-noise ratio due to the decrease of the signal with time, nor by forward-modeling the effect of time-dependent flows on the travel times. Second, the potential relevance of a $1/T^2$ term in the noise model, which may arise in addition to the usual $1/T$ dependence of the variance \citep[see Eqs. (13) and (B.6) in][]{Fournier2014}, may be studied {analytically}. Third, this study should be extended to spherical geometry and to include different types of flow fields, including simulations with different spectral shapes for the flow field \cite[e.g.,][]{Cossette2016,Featherstone2016} {and different flow topologies {\citep[e.g.,][]{Spruit1997,Brandenburg2016,Bekki2017b,Hotta2019,Anders2019}}}.

On the one hand, the results presented here give evidence that the method of \citetalias{Hanasoge2012} correctly estimates the order of magnitude of the flow under the assumption that solar flows were similar to MURaM (if consistent normalizations of spherical harmonics are used, see also \citeauthor{BirchRevisitingInPrep}, in prep.). However, the observed spectrum then leads to the conclusion that the convective conundrum persists at depth. If the vertical correlation {function of flow components} in the Sun is very different to MURaM, for example, with a very different correlation length, this conclusion may well be different. {Determining the correlation function of solar convection may be simpler if a parameterized version of this function were available, for example, obtained by comparing different simulations.}

On the other hand, we cannot resolve the inconsistency of these results with those obtained by \cite{Greer2015} {using ring-diagram analysis} at this stage. A deeper analysis of that method is required to resolve this issue, see also \cite{Nagashima2020}. We are confident that by a careful and detailed evaluation of all aspects of both methods this discrepancy can be understood and resolved. {In addition, global helioseismic methods might soon be used to infer the spectrum of solar convection at depth \citep[e.g.,][]{Roth2003,Woodard2016,Mani2020}. Helioseismology of convection is thus becoming a promising topic to be studied in the near future.}

\begin{acknowledgements}
We thank Matthias Rempel for providing the MURaM simulations. VB thanks {the referee for valuable feedback and Robert Cameron}, Markus Roth, and Shravan Hanasoge for valuable discussions. This work was supported in part by the Max Planck Society through a grant on PLATO Science. The computational resources were provided by the German Data Center for SDO through a grant from the German Aerospace Center (DLR). We acknowledge partial support from the European Research Council Synergy Grant WHOLE SUN \#810218. This work used NumPy \citep{Oliphant2006,vanderWalt2011}, matplotlib \citep{Hunter2007}, and SciPy \citep{Virtanen2020}.
\end{acknowledgements}



\appendix

\section{Detailed formulas and derivations}
\label{appGB04FFT}

\subsection{Fourier transform convention}

For the forward and inverse discrete Fourier transforms of 2D spatial variables, we used the convention \citep[see also][]{GB2004}
\begin{align}
\hat X(\bk) = X(\bk) &= \frac{h_x^2}{(2\pi)^2} \sum_{\bx} X(\bx) \, e^{-i \bk \cdot \bx}, \\
X(\bx) &= h_k^2 \sum_{\bk} X(\bk) \, e^{i \bk \cdot \bx},
\end{align}
{where $\bk=(k_x,k_y)$ is the Fourier variable (wave vector), $\bx=(x,y)$ is the horizontal spatial variable,} $h_x=L/N_x$ is the sampling distance in $x$, $h_k=2\pi/L$ {is} the resolution in Fourier space, $L$ the spatial extent of the data in {one} direction, and $N_x$ is the number of points in dimension. We used the same symbol for a quantity and its Fourier transform and indicate the Fourier transform by the use of the Fourier variable.

Then, the Fourier transform of the convolution
\begin{align}
(X * Y)(\bx) &= h_x^2 \sum_{\bx'} X(\bx') Y(\bx - \bx')
\end{align}
is
\begin{align}
\widehat{(X * Y)} (\bk) &= (2\pi)^2  X(\bk) Y(\bk).
\end{align}
Parseval's theorem becomes, with $Y(x)=X(-x)$
\begin{align}
{\overline{X(\bx)^2}} &=\frac{1}{N} \sum_{\bx} X(\bx)^2
= h_k^4 \sum_{\bk} | X(\bk)|^2.
\end{align}

\subsection{Travel times}
For the noiseless forward-modeled travel times, this means
\begin{align}
\tauNR(\bx) &= \sum_z h_z \, \Bigg( h_x^2  \sum_{\bx'} \bu(\bx',z_m) \cdot \bK(\bx-\bx',z)  \Bigg),  \\
\tauNR(\bk) &=  \sum_z h_z \,\Bigg( (2\pi)^2  \bu(\bk,z) \cdot \bK(\bk,z_m) \Bigg) .
\end{align}

\subsection{Covariance and noise}

The noise covariance matrix is defined as
\begin{align}
 \Lambda(\bx - \bx') &= \langle\epsilon(\bx) \, \epsilon(\bx')\rangle. \label{eqCovDef}
\end{align}
In order to generate realizations of the noise, we set
\begin{align}
\epsilon(\bx) &= h_k^2 \sum_{\bk} \Big(\mathcal{N}(\bk) \, \sqrt{\Lambda (\bk)} \, C \Big)\, e^{i \bk \cdot \bx} \label{eqNoiseDef},
\end{align}
or equivalently,
\begin{align}
\epsilon(\bk) &= \mathcal{N}(\bk) \, \sqrt{\Lambda (\bk)} \, C \label{eqNoiseDefk}      
\end{align}
with
\begin{align}
C&= \frac{1}{h_k} = \frac{L}{2\pi} \label{eqNoiseDefConst},
\end{align}
where $\mathcal{N}(\bk)$ are complex Gaussians with independent real and imaginary parts and unit variance that are independent for different \textit{\textbf{k}}$\bk$, except for {the condition} $\mathcal{N}(-\bk)=\mathcal{N}(\bk)^*$.

Then, we can easily verify that Eq.~\eqref{eqCovDef} is verified,
\begin{align}
\langle\epsilon(\bx) \, \epsilon(\bx')\rangle &= h_k^2 \sum_{\bk} \Big(\sqrt{\Lambda (\bk)} \, C \Big)\, e^{i \bk \cdot \bx} \nonumber \\
&\quad \times h_k^2 \sum_{\bk'} \Big( \sqrt{\Lambda (\bk')} \, C \Big)\, e^{i \bk' \cdot \bx'} \underbrace{\langle \mathcal{N}(\bk) \, \mathcal{N}(\bk') \rangle}_{\delta_{\bk,-\bk'}}\\
&= h_k^4 C^2  \sum_{\bk}  \Lambda (\bk)\, e^{i \bk \cdot (\bx-\bx')}= h_k^2 C^2  \Lambda (\bx-\bx') \\
&= \Lambda (\bx-\bx').
\end{align}
We {use} $\sigma^2=\langle\epsilon(\bx)^2\rangle = \Lambda (\bx-\bx'=0)$.

\subsubsection{Relation to Eq. (33) in \cite{GB2004}}

As a consequence of the above definitions, we have
\begin{align}
\langle |\epsilon(\bk)|^2 \rangle&= \frac{h_x^4}{(2\pi)^4} \sum_{\bx,\bx'} \underbrace{\langle \epsilon(\bx) \epsilon(\bx')\rangle}_{=\Lambda(\bx-\bx')=\Lambda(-\by)} \underbrace{e^{i\bk\cdot\bx - i\bk\bx'}}_{=e^{-i\bk \by}} \\
&= \frac{h_x^2}{(2\pi)^2} \sum_{\bx} \Lambda(\bk) = \frac{L^2}{(2\pi)^2}\Lambda(\bk) \\
&= \frac{1}{h_k^2} \Lambda(\bk) \label{eqNoiseCovRelationshipGB04}\\
\Rightarrow \Lambda(\bk) &= h_k^2 \langle |\epsilon(\bk)|^2 \rangle \\
\Rightarrow \Lambda(\bx-\bx') &= h_k^2 \sum_\bk \left( h_k^2 \langle |\epsilon(\bk)|^2 \rangle  \right) e^{i\bk\cdot(\bx-\bx')},
\end{align}
which differs by a factor of $h_k^2$ from Eq. (33) in \cite{GB2004}.

\subsubsection{Continuous setting}

In a continuous setting \citep{GB2002},
\begin{align}
\langle\epsilon_{j}(\bk) \, \epsilon_{j'}(\bk')\rangle = \delta_D(\bk-\bk') \Lambda_{jj'}(\bk).
\end{align}
To transform back to a discrete setting, we can substitute
\begin{align}
\delta_D(\bk-\bk') \leftrightarrow \frac{1}{h_k^2} \delta_{\bk,\bk'}
\end{align}
and we also obtain $C=1/h_k$.

\subsection{Power spectra and equivalent mode amplitudes in spherical harmonics}
\label{appSHT}

Following \citeauthor{BirchRevisitingInPrep} (in prep.), we converted the power spectrum of any quantity from Fourier space into an equivalent power spectrum in spherical harmonic space with the same variance at a single location or per area. As our Fourier transform definition is differently normalized than in \citeauthor{BirchRevisitingInPrep} (in prep.), we here summarize all formulas appropriate for the convention chosen here.

For any quantity $q$ {(e.g., $q=\tau$ or $q=u_x$)}, we write
\begin{align}
{\overline{q(\bx)^2}} &= 
\frac{1}{N} \sum_{\bx} q(\bx)^2 = h_k^4 \sum_\bk |q(\bk)|^2
= h_k r \sum_{k\in I} \frac{\PFFT_q(k)}{h_k r},  \\
\PFFT_q(k) &= \sum_{\bk \in I_{kr}} h_k^4 |q(\bk)|^2,
\end{align}
{where $h_k r$ is the equivalent resolution in harmonic degree, $I=\{j\,h_k|j=0,\ldots,N_x-1\}$, the sets $I_\ell$ are defined in Section~\ref{secCalCurve}, and we use $\ell=k r$ to convert wave number into spherical harmonic degree.} We then define the amplitude per mode by
\begin{align}
q^{\rm{fft}}(\ell) &= \sqrt{\frac{\PFFT_q(\ell/r)}{(2\ell+1)h_k r}}\\
&= \sqrt{\frac{\sum_{\bk \in I_\ell} h_k^3 |q(\bk)|^2}{(2\ell+1)r,}}
\end{align}
so that
\begin{align}
{\overline{q(\bx)^2}} &= h_k r \sum_{k \in I} (2kr+1) \, q^{\rm{fft}}(kr)^2 \\
&\approx \sum_{\ell=0}^{\ell_{\rm{max}}} (2\ell+1) \, q(\ell)^2,
\end{align}
where $ q(\ell)^2$ is an interpolated power per spherical harmonic mode,
\begin{align}
q(\ell)^2 &=  \frac{P_q(\ell)}{2\ell+1}, \\
P_q(\ell) &= \frac{\PFFT_{q,i}(\ell/r)}{h_k r},
\end{align}
where $\PFFT_{q,i}$ is an interpolated version of $\PFFT_{q}$. As we work in Cartesian geometry{, this is equivalent to assuming that the radial coordinate $r$ is approximately constant throughout the domain. We thus }assume $r=\rm{const}=R_\odot$. For $u_x$ at $r=0.96R_\odot$, the difference due to that {compared to assuming $r=\rm{const}=0.96 R_\odot$} is only $2\%\approx 1-\sqrt{0.96}$.

The corresponding energy spectrum $E_q$ as defined by \citeauthor{BirchRevisitingInPrep} (in prep.) satisfies
\begin{align}
\frac{r}{2} {\overline{q(\bx)^2}}
&=  h_k r \sum_{k \in I} \frac{r}{2} (2kr+1) \, q^{\rm{fft}}(kr)^2 \\
&=  h_k r \sum_{k \in I} E_q(kr) \approx   \sum_{\ell=0}^{\ell_{\rm{max}}} E_q(\ell), \\
E_q^{\rm{fft}}(\ell)&= \frac{r}{2} (2\ell+1) \, q^{\rm{fft}}(\ell)^2, \\
E_q(\ell)&= \frac{r}{2} (2\ell+1) \, q(\ell)^2.
\end{align}

{In summary, the quantity $E_q$ is directly comparable to $E_\phi$ from \citeauthor{BirchRevisitingInPrep} (in prep.) when we here choose $q=u_x$.}

\section{Details on travel-time geometry}
\label{appGeometry}

\begin{table}
        \caption{\label{tabDeepFocusGeom}Details of the deep-focus travel-time geometry.}
        \centering
        \begin{tabular}{lccccccr}
                \hline\hline
                \vspace{0.5mm}
\# & $i_1$ &  $i_2$ &  $\Delta_1$ &  $\Delta_2$ &  $\Delta$ & $t_{\rm{inner}}$  &  $t_{\rm{outer}}$ \\
&       &        &  $[\deg]$   &  $[\deg]$   &  $[\deg]$ & $[\rm{min}]$      &     $[\rm{min}]$  \\
\hline
0 &  8  &  8 &  $3.28$  &  $3.28$  &  $6.56$ &  $39.75$ & $58.50$ \\
1 &  9  &  7 &  $3.75$  &  $2.81$  &  $6.56$ &  $39.75$ & $58.50$ \\
2 &  10  &  6 &  $4.22$  &  $2.34$  &  $6.56$ &  $39.75$ & $58.50$ \\
3 &  11  &  6 &  $4.69$  &  $2.34$  &  $7.03$ &  $41.25$ & $60.00$ \\
4 &  12  &  6 &  $5.16$  &  $2.34$  &  $7.50$ &  $42.00$ & $60.75$ \\
5 &  13  &  5 &  $5.62$  &  $1.88$  &  $7.50$ &  $42.00$ & $60.75$ \\
6 &  14  &  5 &  $6.09$  &  $1.88$  &  $7.97$ &  $43.50$ & $62.25$ \\
7 &  15  &  5 &  $6.56$  &  $1.88$  &  $8.44$ &  $44.25$ & $63.00$ \\
8 &  16  &  5 &  $7.03$  &  $1.88$  &  $8.91$ &  $45.75$ & $64.50$ \\
9 &  17  &  5 &  $7.50$  &  $1.88$  &  $9.38$ &  $46.50$ & $65.25$ \\
10 &  18  &  4 &  $7.97$  &  $1.41$  &  $9.38$ &  $46.50$ & $65.25$ \\
11 &  19  &  4 &  $8.44$  &  $1.41$  &  $9.84$ &  $48.00$ & $66.75$ \\
12 &  20  &  4 &  $8.91$  &  $1.41$  &  $10.31$ &  $48.75$ & $67.50$ \\
13 &  21  &  4 &  $9.38$  &  $1.41$  &  $10.78$ &  $49.50$ & $68.25$ \\
                \hline
                \hline
        \end{tabular}
        \tablefoot{The total travel distance is $\Delta=\Delta_1+\Delta_2$. The exact distances of the arcs to the central point were computed by $\Delta_j = (i_j-1) \,\Delta x$, where $\Delta x=0.46875^\circ$ is the spatial resolution of the data. The quantities $t_{\rm{inner}}$ and $t_{\rm{outer}}$ define the window that we used to model the travel-time fit \citep{GB2004}.}
\end{table}

To compute the kernels, we used the data analysis filters and the geometrical setup of \citetalias{Hanasoge2012}, which are summarized as follows. The data were filtered using a high-pass filter that includes a raised cosine,
\begin{align}
        f_1(\nu) &= 
        \begin{cases}
        0 & \text{for} \,\nu < 1.1\,\rm{mHz}, \\
        \cos\left(\frac{\pi(|\nu|-1.5\,\rm{mHz})}{0.4\,\rm{mHz}}\right) & \text{for}\, 1.1\,\rm{mHz} \leq \nu \leq 1.5\,\rm{mHz},\\
        1 & \text{for}\, 1.5\,\rm{mHz} < \nu \leq 5.0\,\rm{mHz}, \\
        0 & \text{for}\, \nu > 5.0\,\rm{mHz},
        \end{cases}  
\end{align}
and a wide phase-speed filter that targets $r=0.96\,R_\sun$,
\begin{align}
f_2(\nu,\ell) &=
\begin{cases}
0 & \text{for} \,\ell=0, \\
\exp\left( \log(0.5) \frac{(|\nu/\ell| - \nu_0)^2}{\rm{HWHM}^2}\right) & \text{for}\, 0<\ell \leq \ell_{\rm max},\\
0 & \text{for}\, \ell > \ell_{\rm max},
\end{cases}
\end{align}
where $\nu_0=20\,\rm{\mu Hz}$, $\rm{HWHM}=6.7\rm{\mu Hz}$, $\ell_{\rm max}=383$, and the filter is applied by multiplying the power spectrum by $f_1 f_2$, or equivalently, by multiplying the Fourier-transformed data by $\sqrt{f_1 f_2}$.

From the filtered data, {\citetalias{Hanasoge2012} measured} travel-time differences in an arc-to-arc way. {This was done by obtaining cross-correlation functions for point-to-point measurements, and then averaging the cross-correlation} functions over arcs before fitting the travel time. Here, {they used} only opposite points on the arcs. Afterward, {\citetalias{Hanasoge2012} averaged} the travel times with a simple arithmetic mean over the different arcs. The arcs are on concentric circles, have a width of 90~degrees, and are in the east and west quadrants. {Table~\ref{tabDeepFocusGeom} gives the distances of the arcs to the central point}. In the case of $\Delta_1=\Delta_2$, the setup is symmetric about the central point and only one pair of arcs exists. Otherwise, both pairs of arcs, one with $\Delta_1$ to the east and $\Delta_2$ to the west of the central point, and the other vice versa, were used, see Figure~2 in \citetalias{Hanasoge2012}.

 \citetalias{Hanasoge2012} projected the data using Postel's projection, centered on the central point between the arcs. {To keep the computational burden} for the kernels and the noise covariance {manageable}, {we did not use Postel's projection, but} approximated the number of points on each arc by $N_{\rm{arc}} = \floor({\frac{\pi}{2}\sin(\max(\Delta_1,\Delta_2)) \, (\Delta x)^{-1}})  + 1$, where we used the floor function. The spatial resolution of the input data is $\Delta x =  0.46875^\circ \times \,\pi / 180^\circ${, which corresponds to $5.69\,\rm{Mm}$ at disc center}.

{\section{More details for the validity of the assumptions}
\label{appMoreDetails}

{In this appendix, we give more detailed background information for our results. Figure~\ref{figExampleUx} shows an example snapshot for $u_x$ taken at the middle of the time period relevant for Fig.~\ref{figExampleTau}. 
\begin{figure}
        \centering

        \includegraphics[width=\linewidth]{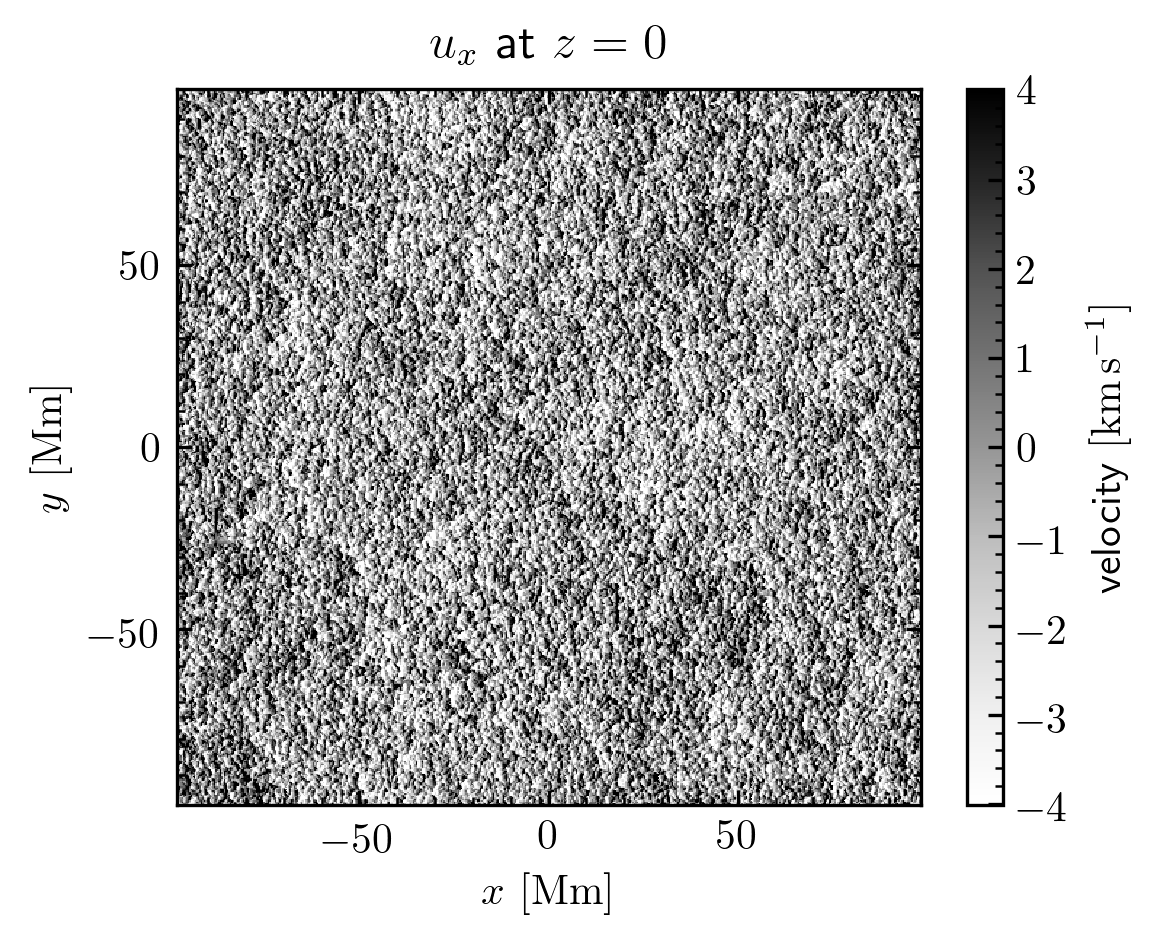}
        
        \caption{{Example realization of the horizontal flow component $u_x$ at the surface ($z=0$). The displayed snapshot is taken in the middle of the time period relevant for the travel times shown in Fig.~\ref{figExampleTau} and is saturated at 33\% of its maximum absolute value. Example realizations for $u_y$ are statistically similar but transposed, and examples for $u_z$ can be found in \citet[][Figs. 2.2 and 2.3]{Lord2014thesis}.}}
        \label{figExampleUx}
\end{figure}
Figure~\ref{figCalCurveDetail} shows a more detailed comparison of the calibration curves $C_l$ and $C_l'/D$, computed for the kernel grid and after interpolating the kernels on the MURaM grid. 
Figure~\ref{figcontribtauk2below099rsun} shows the contribution of the region below $0.99\,R_\sun$ to the travel-time spectrum.

\begin{figure}
        \centering
        \includegraphics[width=\linewidth]{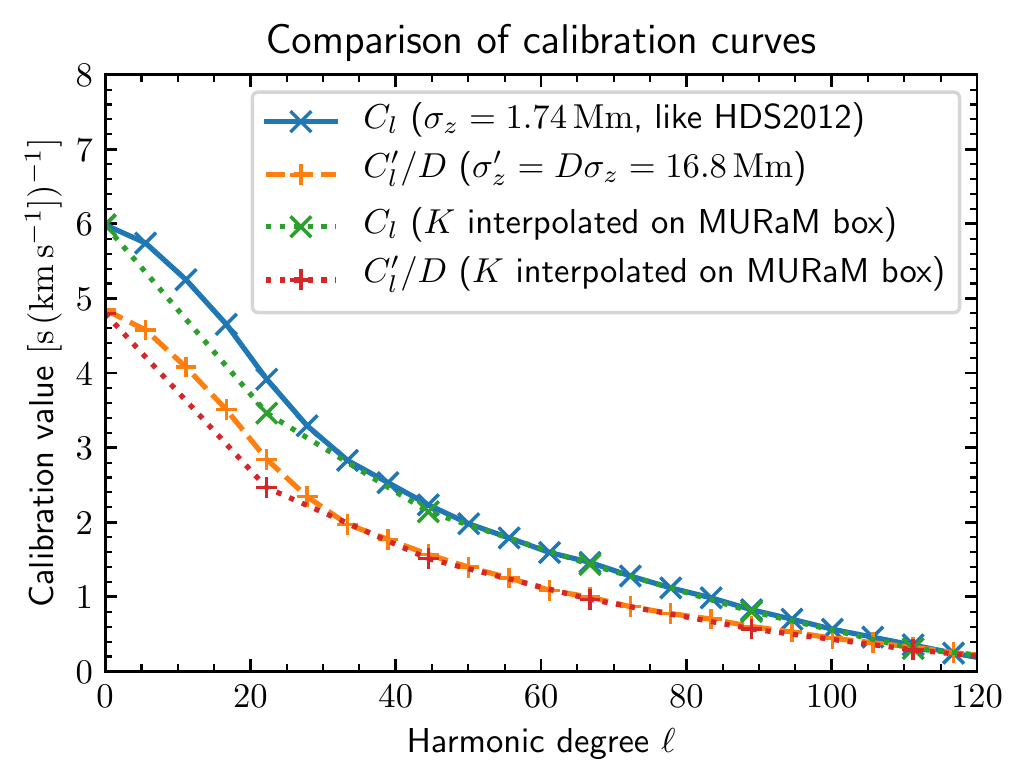}
        \caption{{Comparison of calibration curves in a more detailed way than in Fig.~\ref{figCalCurve}. When the kernels are interpolated onto the MURaM grid, the results do not substantially change (dotted curves), except at $\ell\approx 20$. The crosses and plusses indicate the grid points in harmonic degree, which were obtained by converting horizontal Fourier modes from Cartesian into spherical geometry (see Appendix~\ref{appSHT}). The kernels were computed on a spatial grid with a four times larger spatial extent than the MURaM grid, therefore the resolution in Fourier space, $h_l=r h_k = r \frac{2\pi}{L}$, is four times finer.}}
        \label{figCalCurveDetail}
\end{figure}

\begin{figure}
        \centering
        \includegraphics[width=1\linewidth]{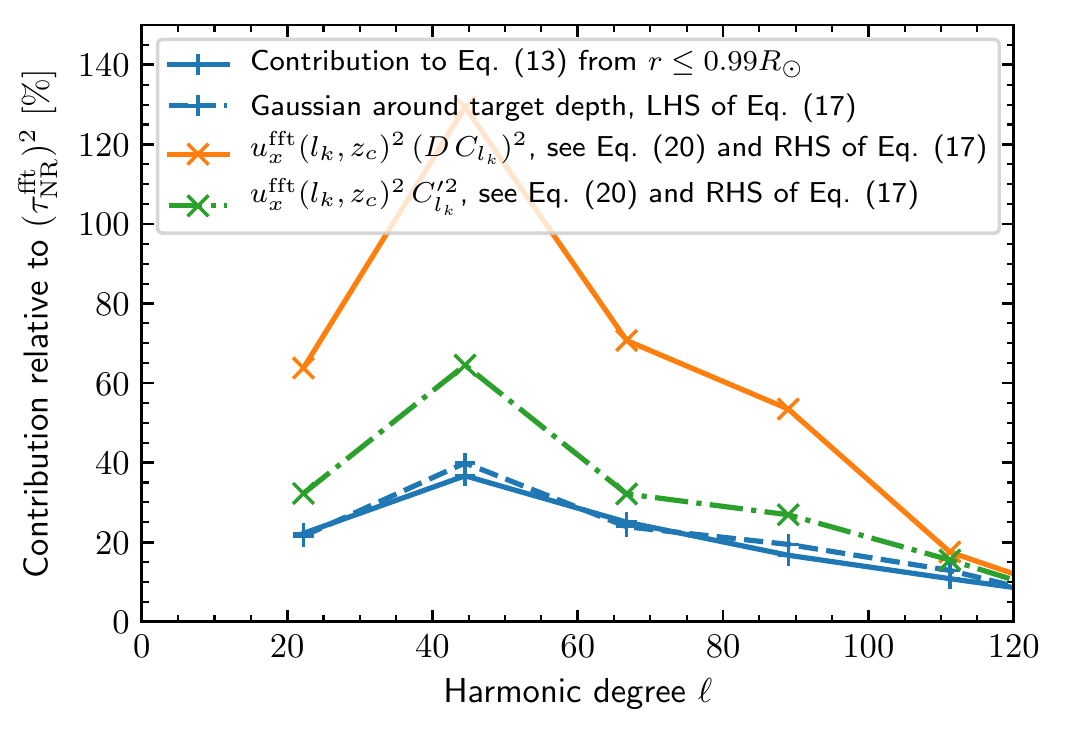}
        \caption{{Contribution to the noiseless travel-time spectrum $(\tauNR^{\rm{fft}})^2$ from below $0.99R_\sun$ (solid blue line) and from a Gaussian-weighted region around the target depth $0.96R_\sun$ (dashed blue) and their {forward models} based on {the actual flow and} two versions of the calibration curve (solid orange and dash-dotted green).}}
        \label{figcontribtauk2below099rsun}
\end{figure}

Finally, we} show a few more detailed plots to verify the assumptions for selected spatial scales $kR_\sun$ and depths $z$. Plots for other spatial scales and depths are qualitatively similar. As a shorthand for Figures~\ref{figKernelkzzprime} - \ref{figKKuu066}, we write for any quantity $Q(\bk)$
\begin{align}
    {\rm{mean}}_{\bk}\, Q(\bk)&= {\rm{mean}}_{|\bk|=k} \, Q(\bk)= \frac{1}{N_{kr}} \sum_{\bk \in I_{kr}} Q(\bk),\label{eqMean}\\
    K_{d}^* K_{d'} \langle u_{d}^* u_{d'} \rangle&= {\rm{mean}}_{|\bk|=k}\, K_{d}^*(\bk,z) K_{d'}(\bk,z') \langle u_{d}^*(\bk,z) u_{d'}(\bk,z') \rangle, \label{eqKKuu}\\
    K_{d}^* K_{d'} &= {\rm{mean}}_{|\bk|=k}\, K_{d}^*(\bk,z) K_{d'}(\bk,z'), \label{eqKK}
\end{align}
and we use $\mathfrak{R}[]$ and $\mathfrak{I}[]$ for the real and imaginary parts of a complex number.
}

\begin{figure*}
        \centering
        \includegraphics[width=1\linewidth]{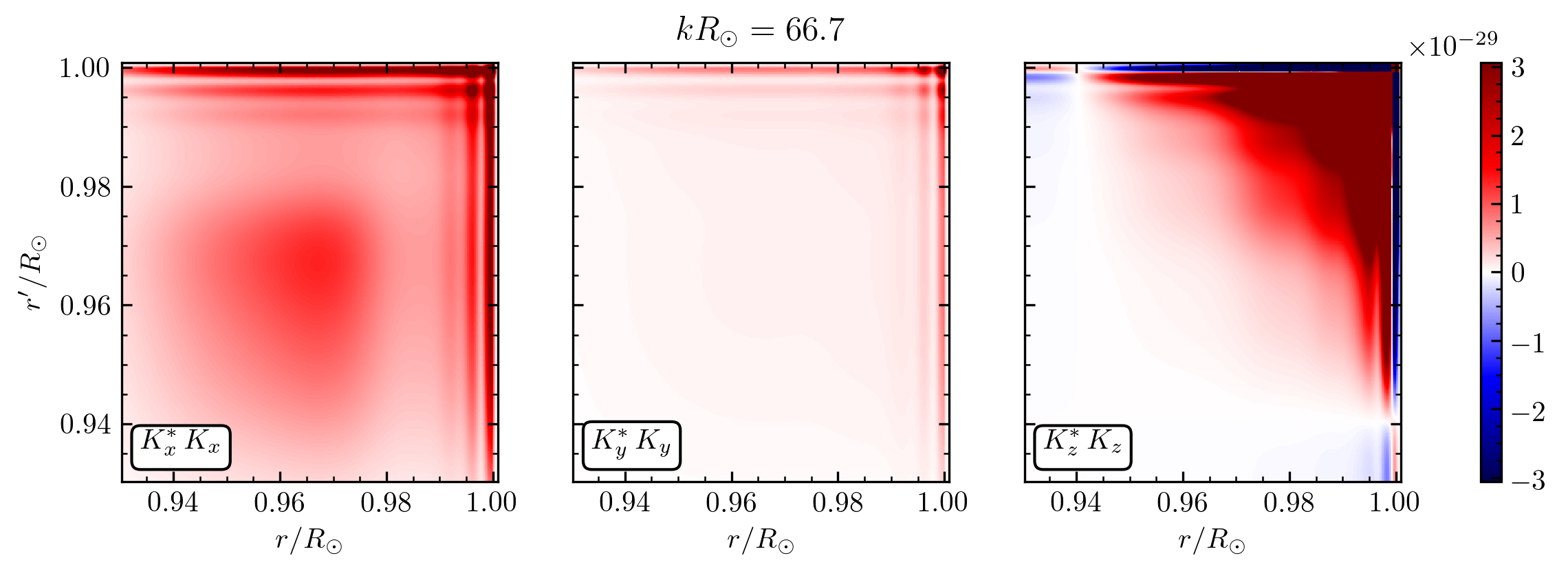}
        \caption{{Azimuthal averages $K_{d}^* K_{d}$ as defined in Eqs.~\eqref{eqMean} and~\eqref{eqKK} saturated at 10\% of the maximum value of the left panel. The units are $\rm{s}^4\,\rm{cm}^{-4}$ (i.e., cgs units). The quantities $K_{d}^* K_{d'}$ are zero for $d \neq d'$, see also Figure~\ref{figKKuuazim066}.}}
        \label{figKernelkzzprime}
\end{figure*}

\begin{figure*}
        \centering
        \includegraphics[width=0.49\linewidth]{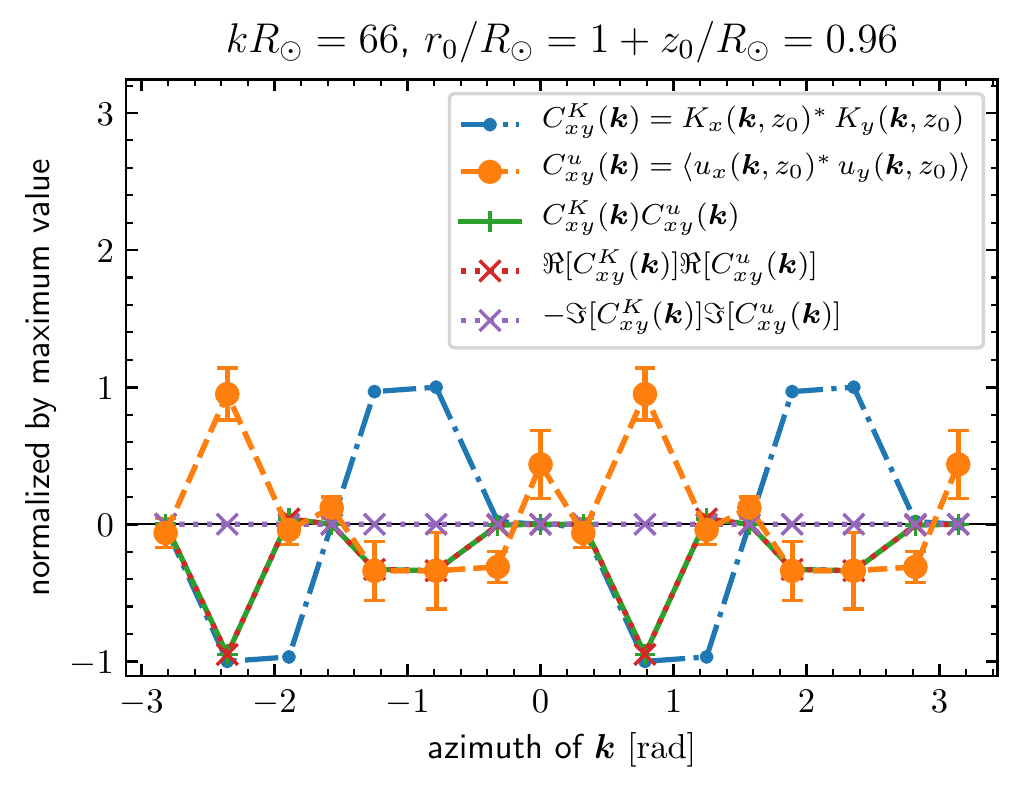}
                \includegraphics[width=0.49\linewidth]{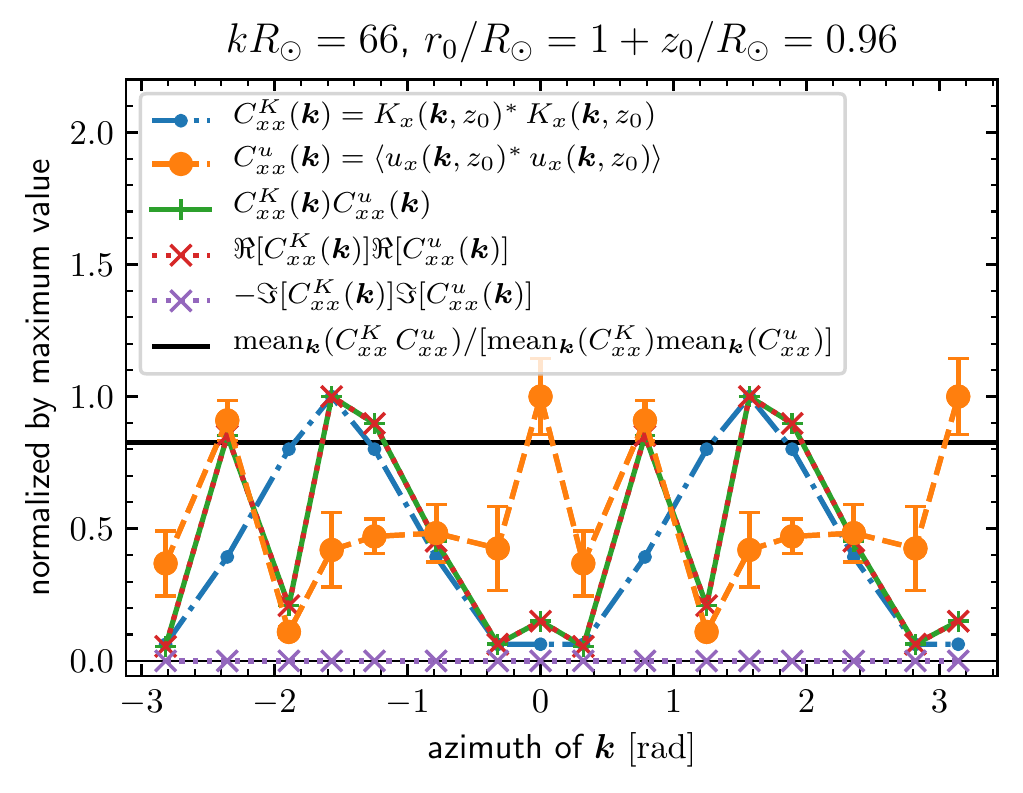}

        \caption{{Exemplary dependence of kernels and flow correlations on the direction of the horizontal Fourier vector.$\bk$ \textit{Left panel}: $\langle u_x^* u_y \rangle$, \textit{right panel}: $\langle u_x^* u_x \rangle$. For the plot, each quantity was divided by its maximum value in the given range. Error bars of $C^{u}$ were obtained from the standard deviation of the 11 available simulation snapshots and are similar for $C^{K} C^{u}$. At other depths and similar wave numbers, the plots are qualitatively similar. In the right panel, we also show that taking the mean before or after multiplying $C^{K}_{xy}$ and $ C^{u}_{xy}$ changes the result (thick solid black line).}}
        \label{figKKuuazim066}
\end{figure*}

\begin{figure*}
        \centering
        \includegraphics[width=1\linewidth]{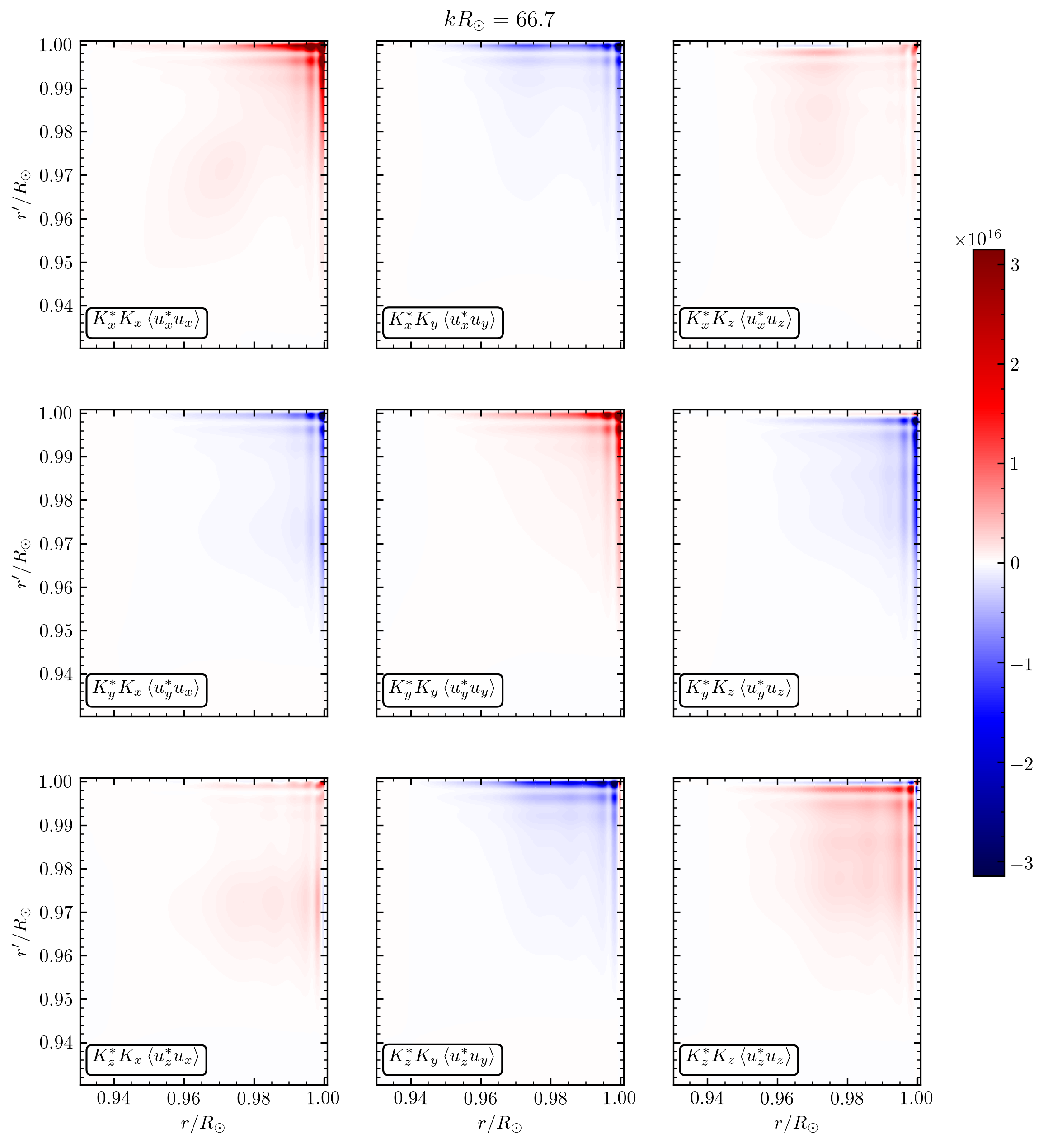}
        \caption{{Azimuthal averages $K_{d}^* K_{d'} \langle u_{d}^* u_{d'} \rangle$ as defined in Eqs.~\eqref{eqMean} and~\eqref{eqKKuu} saturated at 10\% of the maximum value of the top left panel. The units are $\rm{s}^2\,\rm{cm}^{2}$ (i.e., cgs units).}}
        \label{figKKuu066}
\end{figure*}

\end{document}